\documentclass[aps,pra,twocolumn,10pt,floatfix]{revtex4-2}
\usepackage{graphicx,epsfig}
\usepackage{amsmath}
\usepackage{amsfonts}
\usepackage{fancyhdr}
\usepackage{listings}

\usepackage{array}
\usepackage{hyperref}
	\hypersetup{
    colorlinks=true,
    linkcolor=blue,
    filecolor=blue,      
    urlcolor=blue,
    }
\usepackage{listings}
\usepackage{xcolor}
\usepackage{lipsum}
\def \TFMtitle {Modeling Public Opinion Dynamics:\\
The Spiral of silence in clustered homophilic networks}

\def \student {Juan Castillo }

\def \emailaddress {emanuele.cozzo@ub.edu}

\def \advisor {Emanuele Cozzo}

\begin{document}

\pagestyle{fancy}
\lhead{\bf The Spiral of silence in clustered homophilic networks}
\rhead{\student  \& \advisor}

\title{\TFMtitle}
\author{\student}
 
\affiliation{Departament de Física de la Matèria Condensada\\ Facultat de F\'{\i}sica, Universitat de Barcelona, 08028 Barcelona, Spain.}
\author{\advisor}
\email{\emailaddress}
\affiliation{Departament de Física de la Matèria Condensada\\ Facultat de F\'{\i}sica, Universitat de Barcelona, 08028 Barcelona, Spain. \\ Universitat de Barcelona Institute of Complex Systems (UBICS), Universitat de Barcelona, Barcelona, Spain \\ Communication Networks \& Social Change research group (CNSC), Open University of Catalonia, Barcelona, Catalonia, 08018, Spain}
\date{\today}

\begin{abstract}
{\bf Abstract:} Public discourse emerges from the interplay between individuals’ willingness to voice their opinions and the structural features of the social networks in which they are embedded. In this work we investigate how choice homophily and triadic closure shape the emergence of the spiral of silence, the phenomenon whereby minority views are progressively silenced due to fear of isolation. We advance the state of the art in three ways. First, we integrate a realistic network formation model, where homophily and triadic closure co-evolve, with a mean-field model of opinion expression. Second, we perform a bifurcation analysis of the associated Q-learning dynamics, revealing conditions for hysteresis and path dependence in collective expression. Third, we validate our theoretical predictions through Monte Carlo simulations, which highlight the role of finite-size effects and structural noise. Our results show that moderate triadic closure can foster minority expression by reinforcing local cohesion, whereas excessive closure amplifies asymmetries and entrenches majority dominance. These findings provide new insights into how algorithmic reinforcement of clustering in online platforms can either sustain diversity of opinion or accelerate its suppression.
\end{abstract}

\maketitle


\section{Introduction}
Public opinion can be understood, within political science and communication research, as the collective configuration of individual preferences and beliefs on matters of public concern, continuously shaped by media, interpersonal communication, and political context \cite{price1992public}.

Beyond the classical approaches in political science and communication studies, the last few decades have witnessed the rise of a complementary line of research that applies statistical physics methods to the study of public opinion formation. This tradition, often situated at the intersection with computational social science, seeks to model opinion dynamics as collective phenomena emerging from simple interaction rules among individuals, thereby capturing the macroscopic patterns of consensus, polarization, and fragmentation that arise in real societies.

Traditionally, these models have focused on the opinion change (for a review see \cite{SocialDynamics}). Although this approach has been greatly explored and has been able to solve multiple questions, it is often overlooked that, in order to change someone's opinion, an alternative must be voiced. However, social studies show that people are not always willing to express their opinion \cite{Matthes2018,Kalogeropoulos2017}.

One of the most influential theories of public opinion formation was proposed in the 1970s by Elisabeth Noelle-Neumann \cite{Neumann1974, NoelleNeumann2004}. In her theory, Noelle-Neumann sees the fear of isolation as a driving force that leads individuals to silence their opinion if they believe to be in the minority. As more people decide to self-silence, their point of view is then even less represented, causing a spiral where the viewpoint is further silenced. This phenomenon has been observed in face-to-face communication \cite{Scheufele2000}, online environments \cite{Zhao2025}, across cultures and topics\cite{Lee2021} , and has been explored through various computational models \cite{Takeuchi2015, Sohn2019}. However, mathematical modelling of the spiral of silence has only recently gained attention.

In 2020, Gaisbauer, Olbrich, and Banisch proposed a model capable of reproducing the behaviour of the spiral of silence \cite{Gaisbauer2020}. In their approach, the authors consider a general interaction network composed of two groups of agents holding opposing opinions. These agents interact to decide whether or not to express their views. Building on this setup, they develop a mean-field model that can be explored using game-theoretical and dynamical systems tools. This framework allows them to investigate how structural features of the social network affect individuals’ willingness to speak out.

However, it is important to note that in their pursuit of a general and analytically tractable model of the spiral of silence, the authors rely on networks sampled from a stochastic block model \cite{Holland1983}. In doing so, they overlook the fundamental mechanisms that govern the formation and evolution of social networks.  Real-world social ties do not emerge randomly but are shaped by underlying microscopic processes that lead to the concrete, observable structures characteristic of complex networks \cite{Barabasi2016}. Additionally, these underlying mechanism often create network topologies in which mean-field solutions fail to accurately predict the real pattern \cite{Baxter2010}. 

In this work, we address this limitation by incorporating more realistic network formation processes into the model of opinion expression. Specifically, we focus on two well-documented mechanisms: triadic closure, which captures the tendency for individuals to form connections with friends of their friends, and choice homophily, which reflects the preference to interact with others who share similar opinions. The combined action of both processes is believed to produce the observed homophily in networks \cite{kossinets2006empirical}, which is and essential characteristic feature of social networks \cite{mcpherson2001birds}.

We adopt the network generation model proposed by Asikainen et al. \cite{asikainen2020cumulative} in which triadic closure and choice homophily act together to produce a more realistic social network prior to the opinion dynamics. Once the network has reached a stable configuration, we activate the opinion expression mechanisms proposed by Gaisbauer et al. \cite{Gaisbauer2020} on the fixed topology. This approach allows us to explore how structurally emergent features affect the macrosociological patterns of willingness to express opinions. In particular, we focus on the role of triadic closure in amplifying or inhibit the emergence of the spiral of silence in the minority viewpoint compared to a network in which it is not present.

In the following sections, we present the opinion-expression model (section \ref{sec:OpinionExpression}), which includes both a multigroup majority game and an agent learning framework. We then introduce the network generation model (section \ref{sec:NetworkModel}), followed by the mapping of structural parameters from the latter to the former (section \ref{sec:map}). Additionally, we will validate our approach by performing Monte Carlo simulations (section \ref{sec:MonteCarlo}). We conclude with a discussion of the results and their implications (section \ref{sec:conclusion}).
\section{Dynamics of opinion expression}\label{sec:OpinionExpression}
In this section, we present the theoretical framework developed by Gaisbauer \cite{Gaisbauer2020}. This mean field baseline model considers two antagonistic groups and evaluates the conditions under which they choose to express or suppress their views. As agents interact, they respond to their local context leading to emergent collective behaviours.

Importantly, this model assumes that agents conserve their opinion through the dynamics. Moreover, it does not account for the possibility that agents may publicly express an opinion contrary to their own due to peer pressure. Thus, while such behaviours are undoubtedly relevant real-world discourses \cite{Lazarsfeld1944, Kuran1989} and some models exist \cite{SocialDynamics, YE2019371}, they fall outside the scope of the present work.

Building upon this model, we perform a detailed analysis of its dynamical behaviour. Specifically, we solve the system for some extreme cases and investigate the existence of bifurcations and phase transitions in the system, providing deeper insight into the model.

To structure this discussion, we first define the network setting and its defining parameters (subsection \ref{Subsect:Structure}). Then, as in \cite{Gaisbauer2020}, we present the model through two complementary approaches. In the first one, the model is formalized as a multigroup majority game and then we explore its Nash equilibrium (subsection \ref{subsect:Nash}). In the second one, a Q-learning formulation is introduced, which captures the adaptative behaviour of agents over time and its non-deterministic behaviour (subsection \ref{subsect:Q}).  Finally, we explore the bifurcations exist in the Q-learning model and examine the different resulting phases (subsection \ref{sub:bif}).

\subsection{Social structure setting} \label{Subsect:Structure}
The relationships between agents in the society can be modeled through a network, where each node represents an agent and each edge represents a social connection. In our model, each agent belongs to one of two opinion groups, labelled $G_1$ and $G_2$, corresponding to opinions 1 and 2 respectively. The number of nodes in each group is denoted as $N_1$ and $N_2$.

Under the mean field approximation, the structure of the network is not captured by a specific graph but rather the probabilities that connections between agents of different groups are formed. Let $q_{ij} \in [0, 1]$ represent the probability that an edge connects a node from group $G_i$ to a node of group $G_j$. Since we are going to consider undirected graphs, these probabilities are symmetric $q_{ij} = q_{ji}$.

Given these connection probabilities $q_{ij}$, we define the transition probability $T_{ij}$ as the probability that a randomly chosen neighbour of a node in group $G_{i}$ belongs to group $G_j$. We can express these probabilities as the fraction of expected neighbours holding opinion $j$ of a typical node in $G_i$ between the expected total neighbours of that node, which gives:
\begin{equation}
 T_{ij} =
  \begin{cases}
   \frac{N_j q_{ij}}{(N_i -1)q_{ii} + \sum_j N_jq_{ij}},        & \text{if } i \neq j, \\
   \frac{(N_i - 1) q_{ii}}{(N_i -1)q_{ii} + \sum_j N_jq_{ij}},        &i = j.
  \end{cases} \label{eq:Transition}
\end{equation}
where we have used that there are not multi-edges and self connections. 

By definition, these probabilities are normalized:
\begin{equation}
    \sum_{j} T_{ij} = 1
\end{equation}

The matrix $\mathbb{T} = [ T_{ij} ]$ plays a central role in the following opinion-expression dynamics as it measures the effective social exposure that a group has to another. In particular, it defines how frequently agents are going to encounter opinions contrary to their own in their closest neighbourhood and thus it is related to the peer-pressure that is going to affect each group. 
\subsection{Nash equilibrium}\label{subsect:Nash}
To formalize the decision making process behind the opinion expression, in \cite{Gaisbauer2020} a multigroup majority game framework is introduced. In the game, each agent have to decide whether they express their opinion or not. Their election will be influenced by their perception of the opinion landscape of the society, which is going to be formed based on the composition of their neighbourhood. However, only those agents that are willing to express themselves contribute to this perceived landscape. As a result, the expression of a given opinion is not only influenced by the number of agents holding it, but also on the expression of its agents, creating a feedback loop that can amplify or suppress minority opinions as in Neumann's spiral of silence theory.

Additionally, there is another assumption introduced in the model which is that the opinion expression entails a cost $\xi$. This account for the effort that one has to put in order to reply to others in online social networks or join a public discussion about some issue. As a consequence of this cost, a simple majority is not enough to have expression.

\begin{figure}
    \centering
    \includegraphics[width=0.9\linewidth]{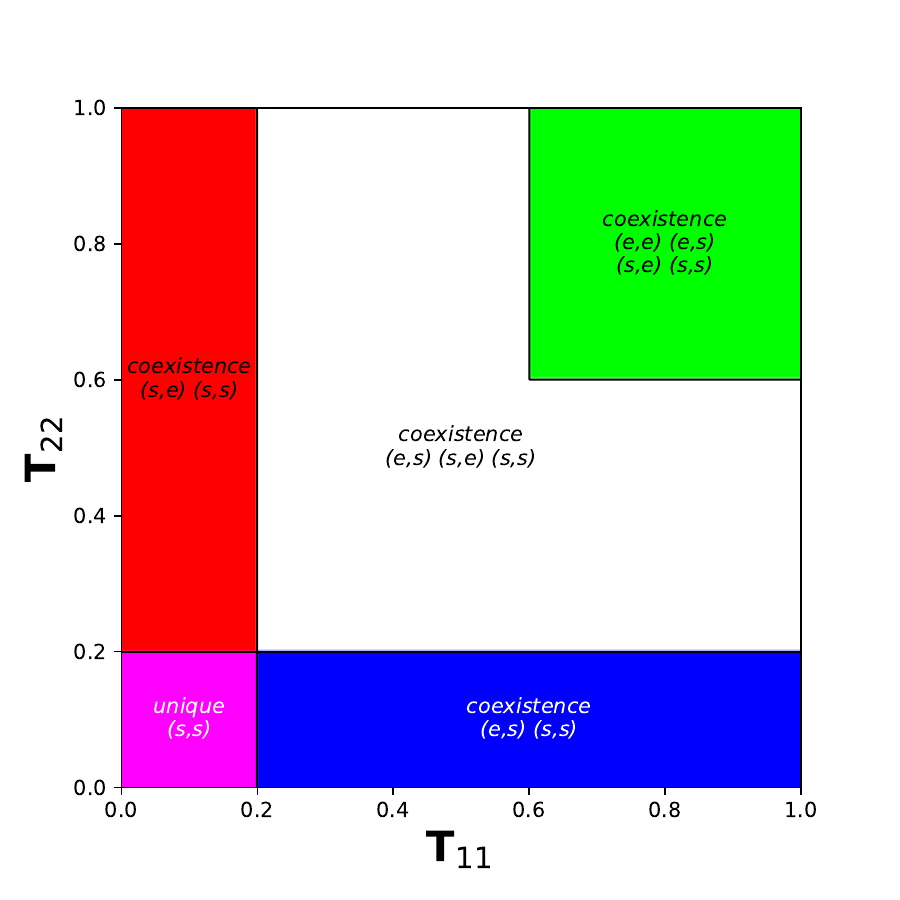}
    \caption{\textbf{Pure-strategy Nash equilibrium regimes.} The equilibria are abbreviated by \textit{e} for opinion expression and by \textit{s} for silence. The first entry in the '(,)' label correspond to the behaviour of $G_1$ and the second for $G_2$. When the structural parameter $T_{ii}$ is smaller that $\xi$ (for reference, $\xi = 0.2$ in the figure), the group is silent. As $T_{ii}$ grows, the group is more willing to express if the other group is silent. When both $T_{11}$ and $T_{22}$ are bigger than $\frac{\xi + 1}{2}$, a new phase is possible where both groups expresses.}
    \label{fig:pure-Nash} 
\end{figure}

Therefore, if we assume fully rational agents and that their decisions are driven only by this mechanism, the preference between expressing or remaining silent must be based on a cost-benefit analysis that can be formalized as a series of Nash equilibria. As the mixed strategies produce metastable states \cite{Gaisbauer2020}, the long-term behaviour of the system can be described in terms of the pure-Nash equilibria. Restricting ourselves to the two opinion group cases, there are only four equilibrium states:
\begin{enumerate}
    \item Both groups express their opinion (which we will denote as (E, E)). The Nash equilibrium inequality is then
    \begin{equation}
 2 T_{11} - 1 -\xi > 0
\label{E-E1}
\end{equation}
\begin{equation}
2 T_{22} - 1 - \xi> 0 
\label{E-E2}
\end{equation}
\item Only group $G_1$ expresses (equilibrium (E, S)). The Nash equilibrium inequality is then:
\begin{equation}
 T_{11}  -\xi > 0
\label{E-S1}
\end{equation}
\begin{equation}
 - 1 + T_{22} - \xi < 0 
\label{E-S2}
\end{equation}
\item Only $G_2$ expresses (equilibrium (S, E)). Similarly to the previous case:
\begin{equation}
  - 1 + T_{11}-\xi < 0
\label{S-E1}
\end{equation}
\begin{equation}
 T_{22} - \xi > 0 
\label{S-E2}
\end{equation}
\item Neither of both groups expresses (equilibrium (S, S)). This is the simplest case as both inequalities reduce to
\begin{equation}
-\xi<0
    \label{S-S}
\end{equation}
\end{enumerate}
The different regimes of pure-strategy Nash equilibrium are summarised in figure \ref{fig:pure-Nash}. Note that, even if the network structure forces a group to be always in a local minority, it can still express itself if the other group remains in silence. However, in order to have expression of all the nodes, it is required that all the nodes see themselves in the majority.

\subsection{Q-learning}\label{subsect:Q}
The game-theoretic approach has important limitations: it cannot determine which equilibrium a given population will reach when multiple Nash equilibria exist, nor can it explain behavioral dynamics when agents are not completely rational. To address these limitations, Gaisbauer et al. \cite{Gaisbauer2020} adopt a reinforcement learning approach based on the formalism proposed by Kianercy et al. \cite{PhysRevE.85.041145}. 

In this dynamical model, actions are selected stochastically according to individual agent preferences. The probability of each action for each agent is parametrized by a Q-function. These probabilities follows then a Boltzmann distribution where the Q-functions play the role of energy. If we denote the probability of agent $i$ expressing as $p_e^{(i)}$, the definition for the expression probability is:
\begin{equation}
    p_{e}^{(i)} = \frac{1}{1+ e^{-\beta Q^{(i)}}}
\end{equation}

The Q-learning model proposed has a random reward determined by with whom the agent interacts. At each time step, when the Q-functions of the agents are updated in order to reinforce actions that have a higher reward. The updating step goes as follows:
\begin{enumerate}
    \item A random agent is selected for expression. It receives a reward $-\xi$ for the cost of expression.
    \item A neighbour of the expressive agent is sampled and talks with its corresponding probability.
    \item If the neighbour also decides to speak and shares the same opinion as the initial agent, an additional reward of $+1$ is received. If the neighbour expresses a contrary opinion, the additional reward is $-1$.
\end{enumerate}

Let us denote by $r^{(i)}(t)$ the reward that agent $i$ receives at time $t$ for expressing themselves.  Then, the value of $Q^{(i)}$ is updated according to:
\begin{equation}
    Q^{(i)}(t + 1) = Q^{(i)}(t) + \alpha' \left[r^{(i)} (t) - Q^{(i)}(t)\right]
\end{equation}
where $\alpha'$ is the learning rate.

This equation can be written in the continuous-time limit by changing $t+1$ to $t + dt$ and $\alpha'$ to $\alpha = \alpha '\,dt$. Doing so:
\begin{equation}\frac{d Q^{(i)}}{dt} = \alpha \left[r^{(i)}(t) - Q^{(i)}\right] \label{eq:QUpdate}
\end{equation}

Through a mean-field approximation, it is possible to show that the agents within each group should converge to the same value over time. As a consequence, we can replace the values of $Q^{(i)}$ in each group for their average. This allows to reduce the system dimensionality to two. From now on, $Q_{1}$ and $p_1$ will denote the $Q$ value and the probability of expression for $G_1$ and $Q_2$ and $p_2$ the analogous for $G_2$. The expected rewards for each group then becomes:
\begin{equation}
    \mathbb{E}\left[ r_1\right] = -\xi + T_{11}p_1 - T_{12}p_2
    \label{eq:Exp.Rew.}
\end{equation}
for $i \in G_{1}$ and
\begin{equation}
    \mathbb{E}\left[ r_{2}\right] = -\xi + T_{22}p_2 - T_{21}p_1
    \label{eq:Exp.Rew2.}
\end{equation}
for $i \in G_{2}$.

The reduced two–dimensional dynamical system is now:
\begin{equation}
  \frac{1}{\alpha} \frac{dQ_1}{dt} = T_{11} \frac{1}{1+e^{-\beta Q_1}} - (1-T_{11}) \frac{1}{1+e^{-\beta Q_2}} - Q_1 - \xi \label{eq:QLearning1}
\end{equation}
\begin{equation}
   \frac{1}{\alpha}\frac{dQ_2}{dt} = T_{22} \frac{1}{1+e^{-\beta Q_2}} - (1-T_{22}) \frac{1}{1+e^{-\beta Q_1}} - Q_2 - \xi  \label{eq:QLearning2}
\end{equation}
where we have used that, by normalization, $T_{12} = 1-T_{11}$ and $T_{21} = 1 - T_{22}$.

\subsection{Bifurcations}\label{sub:bif}
In this subsection, we will present original result regarding the dynamical system described in the previous subsection and its bifurcations. The starting point will be the nullclines, that can be easily obtained from equations (\ref{eq:QLearning1}) and (\ref{eq:QLearning2}). For the dynamics of $Q_1$,

\begin{equation}
    \frac{1}{1+e^{-\beta Q_{1}}} = \frac{Q_{1}}{T_{11}} + \left[\frac{\xi}{T_{11}} + \frac{1-T_{11}}{T_{11}}\frac{1}{1+e^{-\beta Q_2}} \right] \label{eq:Nullcline}
\end{equation}
This defines the nullcline as the intersection point between a sigmoid function and a straight line whose y-intercept depends on the $Q_{2}$ value in an increasing dependence. As a consequence of the shape of sigmoid-functions, this is always guaranteed to have at least one solution for every $Q_2$. As the same is true for the $\frac{dQ_2}{dt} = 0$ nullcline, the system has at least one fixed point. 

Equation (\ref{eq:Nullcline}) can have one or three solutions, the latter only being possible if $\beta T_{11} \geq 4$ and if the y-intercept value is in a given interval $\Omega(\beta, T_{11})$, with $\Omega(\beta, T_{11}) \subset [0, 1]$. Whenever there are 3 solutions, there is at least one value of $Q_{1}$ that is positive and another that is negative. As $0\leq \frac{1}{1+e^{-\beta Q_2}} \leq 1$, the  y-intercept is bounded in $\left[\frac{\xi}{T_{11}}, \frac{\xi + 1-T_{11}}{T_{11}}\right]$, which is sometimes contained in $\Omega(\beta, T_{11})$. When this happens, by nullcline geometry, we can assure that there is going to be a stable fixed point in which $G_{1}$ speaks out and another in which $G_{1}$ will remain silent, independently of what decides the group $G_{2}$ to do.

\begin{figure}[!b]
    \centering
    \includegraphics[width=0.9\linewidth]{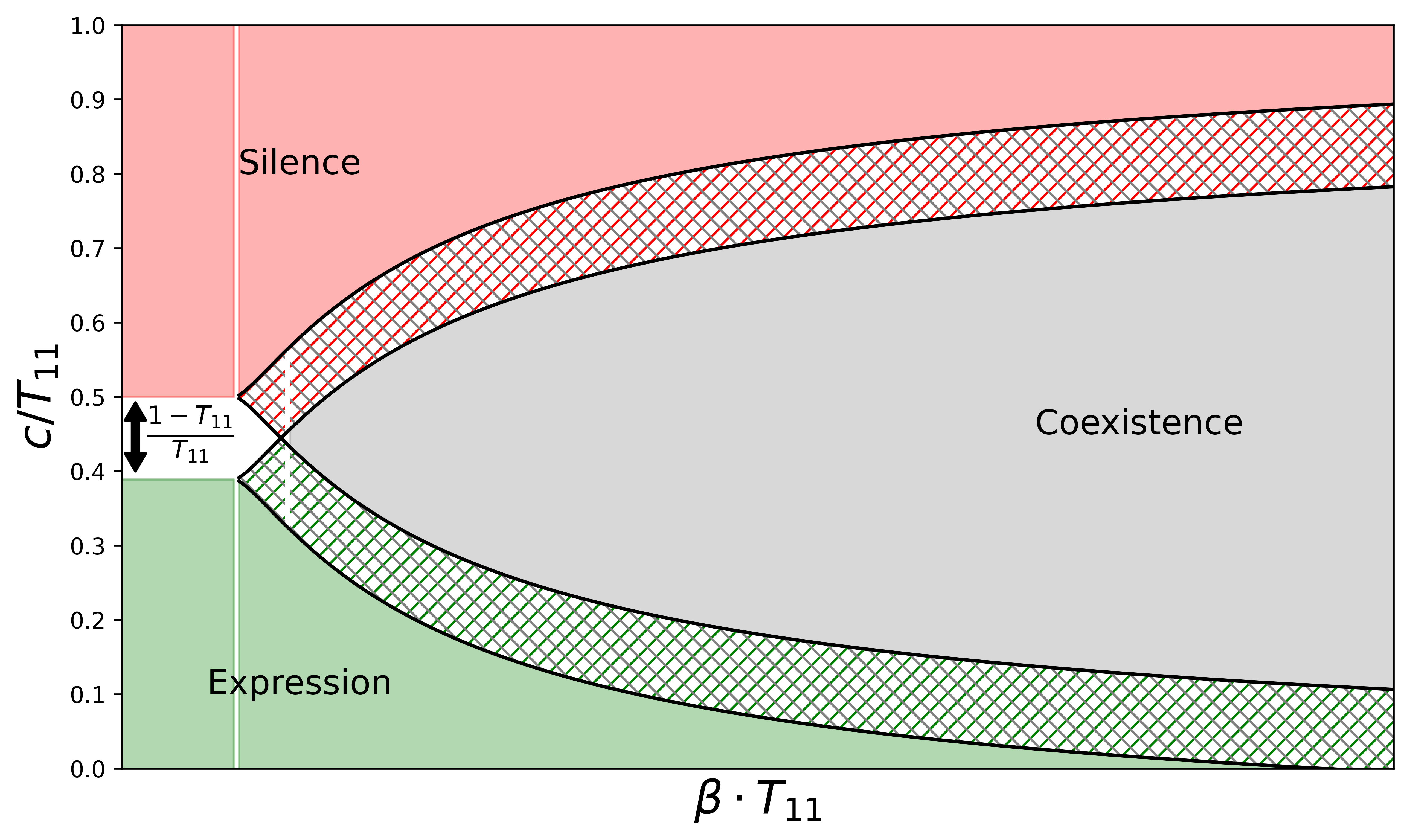}
  \caption{
\textbf{Phase diagram for the expression of one group}. Colored regions represent parameter combinations where the phase of the group is guaranteed, regardless of the behavior of the other group. The hatched region indicates that a specific phase is guaranteed, but coexistence is also possible depending on the state of the other group.
}

    \label{fig:Phase_dia}
\end{figure}
In the same manner, we can ensure that whenever there is only one fixed point, sometimes only an expressive state of the group is possible. For example, if $\xi > T_{11}$, the value of the linear function will always be greater than one whenever $Q_{1}>0$ and therefore the crossing point should be in the negatives $Q_{1}$ $\forall Q_{2}$, leading to silenced $G_1$ independently of the expression of $G_2$. The analysis performed can be summarized in the phase space plotted in figure \ref{fig:Phase_dia}. Note that, since the dynamical equation for $Q_2$ is symmetric, the plot is also valid for the group $G_{2}$.

An important result that can be obtained from figure \ref{fig:Phase_dia} is that, in order to get an expressive state for a group at a finite temperature, the group has to also be able to express at lower temperatures. As the Nash equilibrium is the $T\to 0$ limit of the dynamical system (see Appendix I), if the Nash equilibrium conditions do not allow a group to speak out or an E-E state, those phases will also not be reachable for any temperature.

The previous analysis does not give the strict frontiers between phases. In order to obtain the bifurcation conditions, we must identify when new fixed points appear or disappear in the system. By inspection, we concluded that the new fixed points arise from saddle bifurcations. The condition necessary for a saddle bifurcation is that both nullclines should be tangent:
\begin{equation}
    \left.\frac{dQ_{2}}{dQ_{1}}\right|_{\dot{Q}_1 = 0} = \left.\frac{dQ_{2}}{dQ_{1}}\right|_{\dot{Q}_2 = 0}\label{Bif}
\end{equation}
In order to compute $ \left.\frac{dQ_{2}}{dQ_{1}}\right|_{\dot{Q}_1 = 0} $, we took equation (\ref{eq:Nullcline}) and solve for $\frac{1}{1+e^{-\beta Q_{2}}}$. Then, we simply applied chain rule:
\begin{equation}
   \frac{d}{dQ_1}\left[\frac{1}{1+e^{-\beta Q_2}}\right] = \frac{\beta}{4\cosh^2(\beta Q_2/2)} \left.\frac{dQ_{2}}{dQ_{1}}\right|_{\dot{Q}_1 = 0}
\end{equation}
An analogous procedure with the other nullcline can be done to obtain $ \left.\frac{dQ_{2}}{dQ_{1}}\right|_{\dot{Q}_2 = 0}$ taking into account that 
\begin{equation}
     \left.\frac{dQ_{2}}{dQ_{1}}\right|_{\dot{Q}_2 = 0} =  \left.\frac{dQ_{1}}{dQ_{2}}\right|_{\dot{Q}_2 = 0}^{-1}
\end{equation}

Finally, by substitution in equation (\ref{Bif}), the saddle bifurcation condition is:
\begin{equation}
\begin{split}
    &\frac{T_{11}T_{22}}{(1-T_{11})(1-T_{22})} =\\ &\frac{\beta T_{11}T_{22}}{(4\cosh^2(\beta Q_1^*/2)-T_{11})(4\cosh^2(\beta Q_2^*/2)-T_{22})}
    \end{split}
\end{equation}
where $(Q_1^*, Q_2^*)$ is a solution of equations (\ref{eq:QLearning1}) and (\ref{eq:QLearning2}).

\begin{figure}[b]
    \centering
    \includegraphics[width=0.8\linewidth]{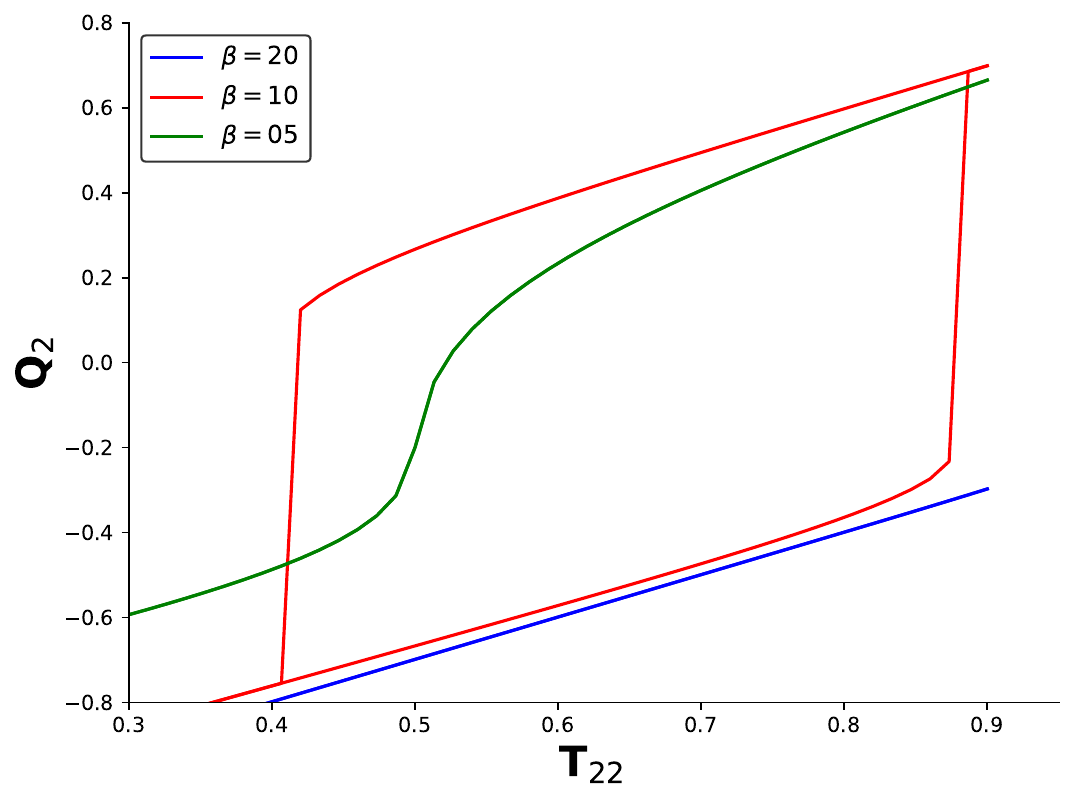}
    \caption{\textbf{Hysteresis cycle in expression dynamics}. As the intra-connectivity grows, new state arise. However, for lower temperatures, the system might remain trapped in previous states, leading to hysteresis. }
    \label{fig:Histeresis}
\end{figure}

As the new points appear through saddle bifurcations, an equilibrated system will not be able to reach the new phase unless the fixed point in which it is equilibrated vanishes. This dynamic gives rise to path dependence and expression hysteresis in the population (figure \ref{fig:Histeresis}). This is especially relevant at low temperatures, as escaping from the co-existence region usually means the silence of one group and due to the hysteresis, the expressive state can not be recovered easily.
\section{The clustered homophylic network}\label{sec:NetworkModel}
In this section, we present the network evolution model proposed by Asikaien et al. \cite{asikainen2020cumulative}. 
Their framework considers two key individual-level mechanisms to simulate the evolution of a social network in societies with different opinions: triadic closure and choice homophily \cite{kossinets2006empirical}. Additionally, the model incorporates some randomness to take into account other possible mechanisms not considered explicitly such as focal closure within shared environments \cite{PhysRevLett.99.228701}. Because these processes reflect fundamental features of social interactions, the generated networks are able to reproduce characteristic observed in real-world communities better than those generated by purely random models.

Choice homophily is defined as people's preference when deciding to whom to connect with once the encounter is produced. It is introduced by creating the edges probabilistically, where these probabilities depend on the group membership of both agents but are homogeneous in any other aspect such as their degree or closeness. On the other hand, triadic closure reflect the tendency of individual's to interact with friends of their friends. Although its importance should not be disregarded in face-to-face relations, it is most relevant in online communities as it is one of the mechanisms used in algorithms to suggest new connections. 

Both processes should not be treated separately, but instead show a cooperative effect. Triadic closure imposes structural constrains to with whom an agent is able to relate. As people tend to be in contact with others that are like-minded, choice homophily creates the circumstances needed to meet other agents with the same opinion through triadic closure. This creates a cycle in which choice homophily and triadic closure act together as an homophily amplifier.

An important aspect to consider is that the network formation model assumes that each agent is aware of the group membership of others, whereas the spiral of silence model explicitly allows agents to not share his or her opinion. To reconcile these two frameworks, we must assume that there exists a non-hidden, observable characteristic that is highly correlated with the expressed opinion in the spiral of silence model.  For instance, one may consider how, in younger generations, gender is often an indicator of the position in the right-left political spectrum \cite{Langsaether2024GenderGap}.

The model is first introduced in subsection \ref{subsection:Agent} as an agent based formulation. Then, in subsection \ref{subsection:MF}, a mean-field approximation is taken to obtain the equations of the dynamics. Finally, in subsection \ref{subsection:structure}, some of the emergent structural features of the model are discussed.

\subsection{Agent based model definition}\label{subsection:Agent}
We begin by describing the agent based model which forms the foundations of the network creation. As in the spiral of silence model, the network is divided into groups with different opinions. Initially, the network is generated randomly with the opinions distributed randomly and independently and with edges created according to some initial probabilities $q^{0}_{ij}$.

From its initial state, the network evolve creating new connection and removing some of the old. At each time step, a node that will act as the focal node is chosen randomly with the condition that it has at least one neighbour. Then, another node of the network is selected as an encounter. This encounter is produced randomly with probability $1-c$ and through triadic closure with probability $c$. If the selected mechanism is triadic closure, then we are going to select a random neighbour of the focal node and then a neighbour of the chosen neighbour. Whatever the mechanism chosen for the encounter is, if the encounter is already connected to the focal node or is the focal itself, then no connection is made.

Once a focal node and an encounter are sampled, we see if the edge between them is created. That is done with probability $s_{ij}$, where $i$ and $j$ are the opinions of the focal node and the encounter respectively. This represent the choice homophile in the society. If the connection is created, then an edge of the focal node is removed as maintaining old relations require some effort and time and thus creating new connections often involve forget the old ones \cite{Saramäki}. This connection is chosen randomly, which is a common choice \cite{Marsili2004} despite than in reality edges that are in a high clustered region of the network are usually stronger \cite{Granovetter1973}. 

The matrix-elements $s_{ij}$ are not necessarily symmetric, as the group $i$ can be more willing to connect with the group $j$ than the other way around. Note also that the choice homophily depends only on the groups both agents belong to.
\begin{figure}
    \centering
    \includegraphics[width=\linewidth]{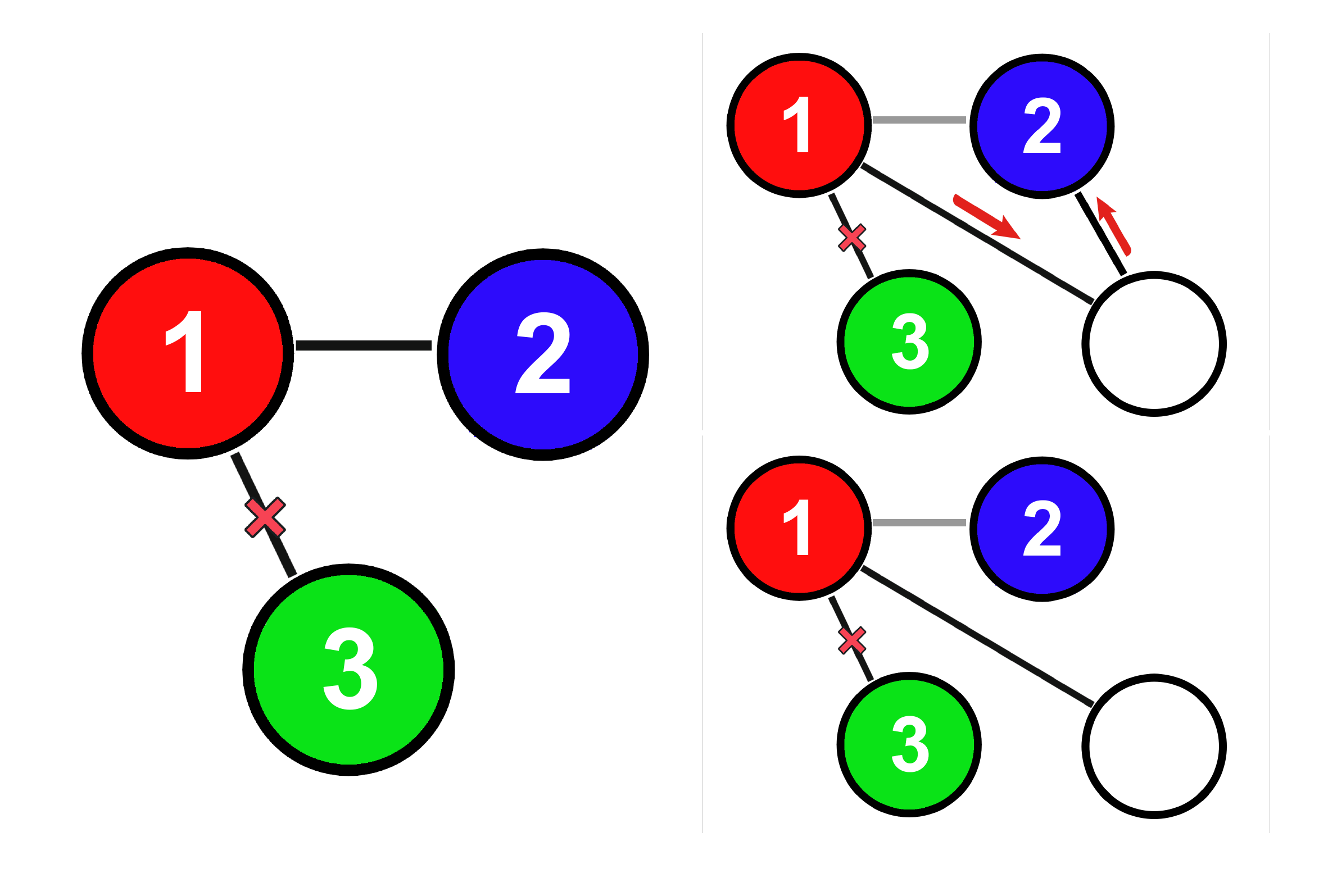}
    \caption{\textbf{Network formation mechanism.} (\textbf{Left}) At each time step an edge is created and another is removed. In this example, an edge between an agent in $G_1$ and another in $G_2$ is created while an edge between the agent in $G_1$ and another agent in $G_3$ is removed. This is a process $1-2-3$. \textbf{(Top right)} The node in $G_1$ connects with the node in $G_2$ through a common neighbour in the process known as triadic closure. (\textbf{Bottom right}) The node in $G_1$ now connects with the node in $G_2$ due to other possible mechanism.}
    \label{fig:Agent Based model}
\end{figure}
As in the dynamic proposed here each time an edge appears another is removed, the total number of links is conserved. Additionally, the degree of the focal node is also conserved in each elemental step. However, the degree distribution is not, as there is a node that has lost one edge and another that has gained one as a consequence of the rewiring. 
\subsection{Mean-field model}\label{subsection:MF}
In this section we will present mean-field evolution equation for the network obtained  in \cite{asikainen2020cumulative}. For that purpose, it is useful to introduce new variables $P_{ij}$ that represent the probability that a given edge connects a node with opinion $i$ and another with opinion $j$. The normalization condition is then written as:
\begin{equation}
    1 = \sum_{i \leq j} P_{ij}
\end{equation}

The elements $T_{ij}$ can be constructed from these probabilities as:
\begin{equation}
    T_{ij} = \begin{cases}
        \frac{P_{ij}}{2P_{ii} + \sum_{k\neq i}P_{ik}}, & \text{if }i\neq j,\\
        \frac{2P_{ii}}{2P_{ii} + \sum_{k \neq i}P_{ik}}, & i = j
    \end{cases}
\end{equation}

The evolution of the matrix elements $T_{ij}$ can be then expressed through the chain rule:
\begin{equation}
    \frac{dT_{ij}}{dt} = \sum_{k<l}\,\frac{\partial T_{ij}}{\partial P_{kl}}\,\frac{dP_{kl}}{dt}
\end{equation}
The evolution of the probabilities $P_{kl}$ on the other hand can be computed form a master equation. In the case of inter-group connections $i\to j$, this master equation takes the following shape:
\begin{equation}
    \frac{dP_{ij}}{dt} = \frac{1}{L}\sum_k\left[P_{i-j-k}+P_{j-i-k} - P_{i-k-j} - P_{j-k-i}\right]
\end{equation}
while in the case of intra-group links $i\to i$:
\begin{equation}
    \frac{dP_{ii}}{dt} = \frac{1}{L}\sum_{j}\left[  P_{i-i-j} - P_{i-j-i}\right]
\end{equation}
where
\begin{equation}
    P_{i-j-k} = n_is_{ij}T_{ik}\left[c(T^2)_{ij} + (1-c)n_j \right]
\end{equation}
are the probability of creating a link from  group $i$ to $j$ while removing a link from group $i$ to $k$ (figure \ref{fig:Agent Based model}).

\subsection{Relevant structural characteristics}\label{subsection:structure}
This model gives rise to a rich phase space. In the case of two groups populations, the model dynamics produce two different social topologies. These topologies can emerge by moving through the phase space in critical Pitchfork bidurcations.

\begin{figure}
    \centering
    \includegraphics[width=\linewidth]{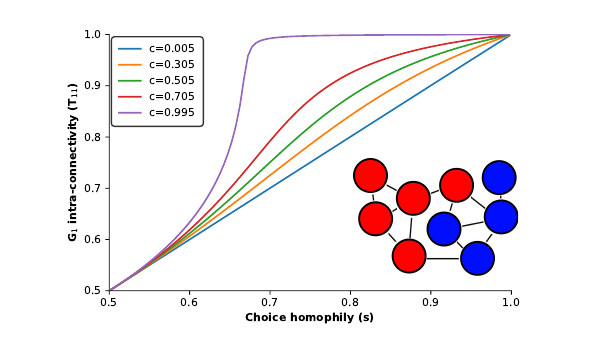}
    \caption{\textbf{Homophilic amplification network.} Predicted fraction of neighbours in the same group $G_{1}$ as a function of choice homophily. Groups have equal size ($n_1= n_2=0.5$) and equal choice homophily ($s_{11}=s_{22} = s$, $s_{12}=s_{21}=1-s$). Inset: example of the typical structure that appears in homophilic amplification fixed-points. Both groups are well connected internally and a only a few inter-groups links are created. }
    \label{fig:Homophilicamplification}
\end{figure}

The simplest case, referred in \cite{asikainen2020cumulative} as homophilic amplification, arises when the initial homophily of both groups is high and agents connect mainly through triadic closure. Due to the structural constrains imposed by these features (figure \ref{fig:Homophilicamplification}), it is much more likely for an agent to have an encounter with another holding the same opinion, and thus, homophilic edges are favoured even in the absence of a high-choice homophily bias. The resulting networks are highly intra-connected within each group and have a low number of intergroup links.

The model is also able to generate core-periphery structures, which are commonly observed in empirical social networks \cite{Rombach2017}. In such cases, one group known as the core becomes densely connected while the other remains in the periphery and it is mostly connected to the core group. In this situation, if the triadic closure probability is high enough, agents in the network are more likely to encounter core members, increasing the probability of heterophilic connections. Although such configurations arise more easily when there is an asymmetry between both groups in group size or choice homophily, they can also emerge spontaneously in symmetric populations if the right conditions are meet.
\begin{figure}
    \centering
    \includegraphics[width=0.8\linewidth]{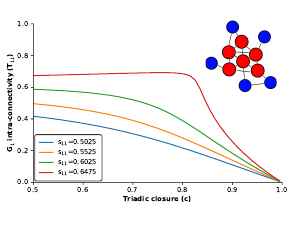}
    \caption{\textbf{Core-periphery network.} Inset: Example of a core-periphery structure. Predicted fraction of neighbours in the same group G1 as a function of triadic closure probability. Groups have equal size ($n_1 = n_2 = 0.5$). $G_2$ has choice homophily $s_{22} = 0.65$ ($s_{12} = 1-s_{11}$, $s_{21}=1-s_{22}$). Note that both red and blue nodes have a high probability of encounter a red node through triadic closure.}
    \label{fig:core}
\end{figure}

Homophily amplification and core-periphery are two competing process that generate structures that are fundamentally different. While the former promotes segregation of the network in structures that resemble echo chambers, the latter introduce a hierarchy that produces an asymmetry between both groups. As a consequence, a small change in the parameters ruling the dynamics might lead to qualitatively distinct social topologies. The system’s sensitivity to initial conditions and structural fluctuations further amplifies this effect, making it possible for nearly identical settings to produce drastically different social configurations.

An additional structural feature that emerges naturally from the dynamics is heavier-tails in the degree distributions. It is well known in network theory that when we are exploring neighbours of another node, it is very likely that those nodes have, on average, more degree that the usual node in the network. Thus, highly connected nodes have higher probabilities to be encountered through triadic closure. This create a 'the rich get richer' dynamics, moving the system from the mean-field description.
\begin{figure*}[t]
\centering
\begin{minipage}[t]{0.75\linewidth}
    \centering
    \includegraphics[width=\linewidth]{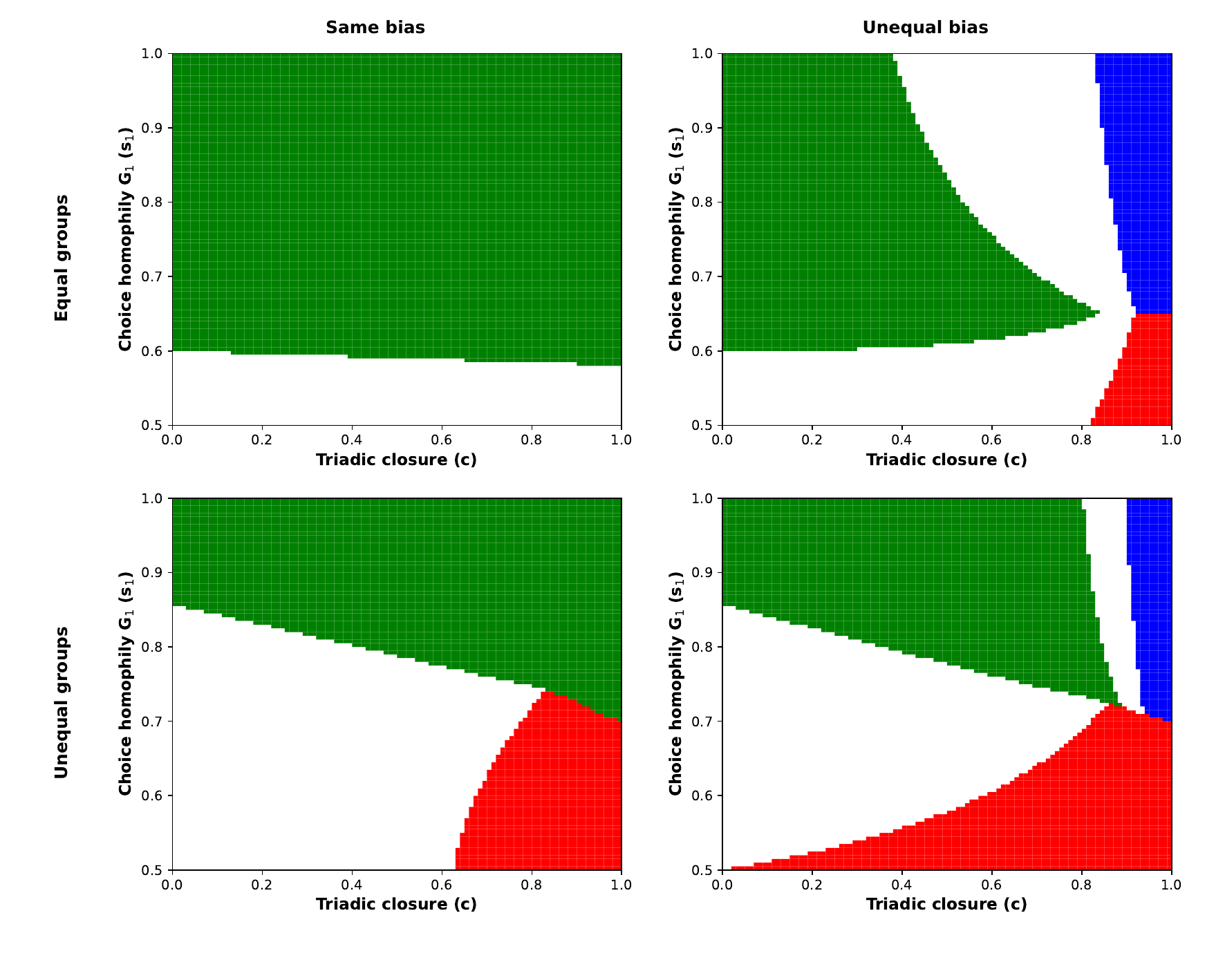}
\end{minipage}%
\hfill
\begin{minipage}[t]{0.23\linewidth}
    \vspace{-3.5cm}
    \includegraphics[width=\linewidth]{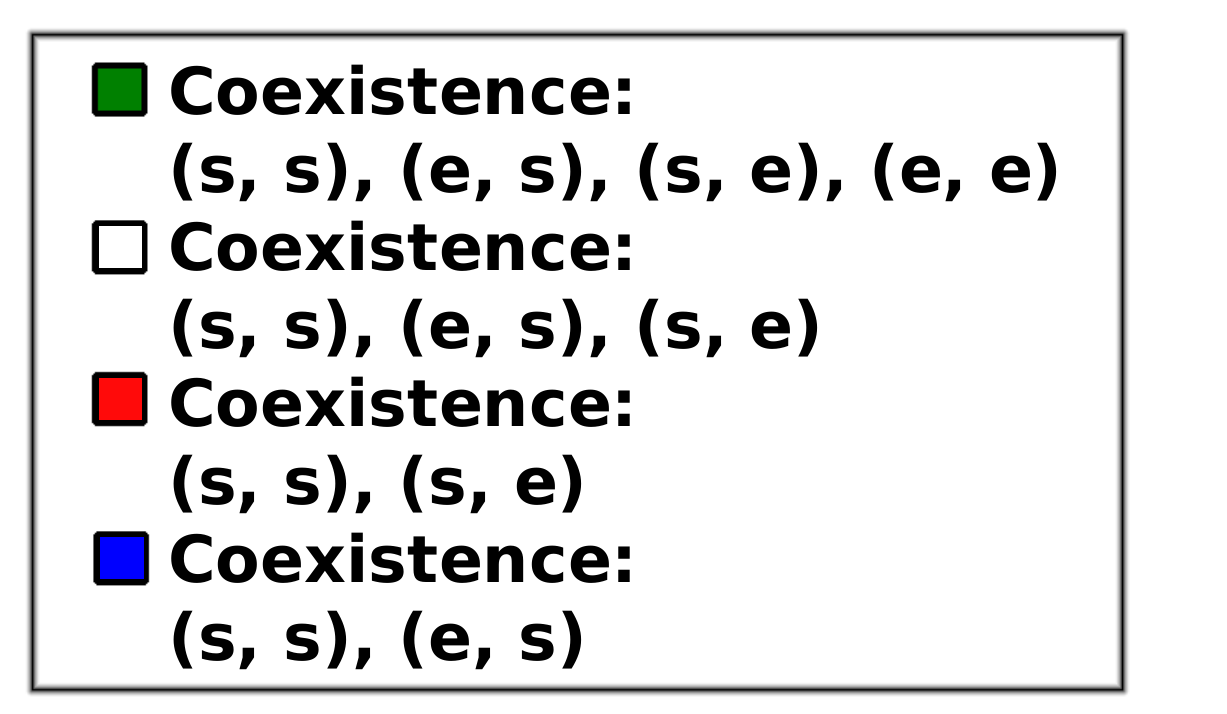}
\end{minipage}
\caption{\textbf{Pure Nash equilibrium in the system.} The colour palette is chosen so it matches the one used in figure \ref{fig:pure-Nash}. The cost is set to $\xi = 0.2$. Each of the planes is divided into a 100x100 grid for the integration. The tolerance for convergence used is $10^{-10}$. (\textbf{Top}) Both opinion groups have the same size ($n_1 = n_2 = 0.5$). (\textbf{Bottom}) Groups are of unequal size ($n_1 = 0.2$, $n_2 = 0.8$). (\textbf{Left}) Choice homophily is the same for both groups ($s_1 = s_2 = s$) and the phase space is explored as a function of $s$ and $c$. (\textbf{Right}) Choice homophily of the majority (group 2) is fixed to $s_2 = 0.65$, and the phase space is explored as a function of $c$ and $s_1$.}
\label{fig:Nasheq}
\end{figure*}

\section{Network-Driven Transitions in Expression Behaviour}\label{sec:map}
In the previous sections, we introduced two different models that form the basis of our analysis: a network evolution model based on triadic closure and choice homophily and an opinion expression model that reproduce Noelle-Neumann's spiral of silence. In this section, our goal is to integrate both frameworks to investigate how key relational processes influences the formation of a public discourse in a society divided in two competing opinions. Due to its central role in the online social networks algorithm, most of this analysis will revolve around the triadic closure mechanism.

In this initial section, this is done through a numerical exploration of the phase space. The starting point is a computational integration of the mean-field equations obtained in subsection \ref{subsection:MF} from a random network without opinion bias (all the $q_{ij}$ elements take the same value, so $T_{11} = n_1$ and $T_{22}=n_2)$. Once the integration has reach the corresponded fixed point, we apply both formalisms introduced in subsection \ref{subsect:Nash} and \ref{subsect:Q} to obtain the mean-field prediction of the opinion dynamics.

By systematically varying all the parameters that rule the network evolution dynamics (group size, choice homophily, and triadic closure probability), we study how the final expression/silence state depends on the behaviour of the individual that conform the network. This approach allow us to identify not only the equilibrium states but also critical transitions and relevant regions in the parameter space. 

As we are considering a network in which there are only two opinion groups, the network dynamics depend on six key parameters: the relative size of $G_1$, denoted by $n_1$ (with $n_2=1-n_1$ by normalization), the triadic closure probability $c$, and the four values of the choice homophily $s_{ij}$. In order to reduce the dimensionality of the parameter space, as in \cite{asikainen2020cumulative}, we are going to consider that $s_{12} = 1-s_{11}$ and that $s_{21} = 1-s_{22}$. These constraints allow for relational asymmetries in both groups while simplifying the analysis.  Additionally, for clarity, we will adopt the convention that group $G_1$ always represents the minority.

The analysis will be divided in two different stages. In subsection \ref{subsection:Nash-eq}, we analyse the system using the Nash equilibrium framework, focusing on the effect of the structural changes by mapping the region of the parameter space to the ideal population behaviour. Later, in subsection \ref{subsection:Q-learning}, the Q-learning framework is recovered to explore the effect of the stochasticity introduced by non-completely rational agents and the path dependence.

\subsection{Influence of Structural Parameters Under Nash Equilibrium}\label{subsection:Nash-eq}

In this subsection, we apply the Nash equilibrium conditions derived in equations (\ref{E-E1}) to (\ref{S-S}) to the matrix elements $T_{ij}$ obtained at the end of the numerical integration. The exploration of the four-dimensional phase space is done through representative two-dimensional planes. Each of these planes is divided into a grid for the numerical integration. The result are shown in figure \ref{fig:Nasheq}.

The first thing to note in the figure is that, regardless of the specific parameter values, at least one of the groups is able to reach an expressive state. This behaviour arises due to the introduction of the choice homophily in the network dynamics, which creates a bias to relate with individuals holding the same opinion. As a consequence, even in the presence of core-periphery structures, this asymmetry produces at least one group which agents are surrounded by supportive neighbourhoods. This reduces the perceived risk of expression, allowing the group to overcome the spiral of silence unless the expression cost is prohibitively high ($\xi >0.5$ is needed at least to produce silence-silence equilibriums). 

Whenever the dynamics of both groups are symmetric (that is when $n_1=0.5$ and $s_{11} = s_{22}$) neither group is favoured over the other. In such plane, networks are usually driven to homophilyc amplification by a cumulative effect of choice homophily and triadic closure.  This symmetry creates favourable conditions for reaching an equilibrium in which both groups are willing to express their opinions.

\begin{figure}[h]
    \centering
    \includegraphics[width=0.8\linewidth]{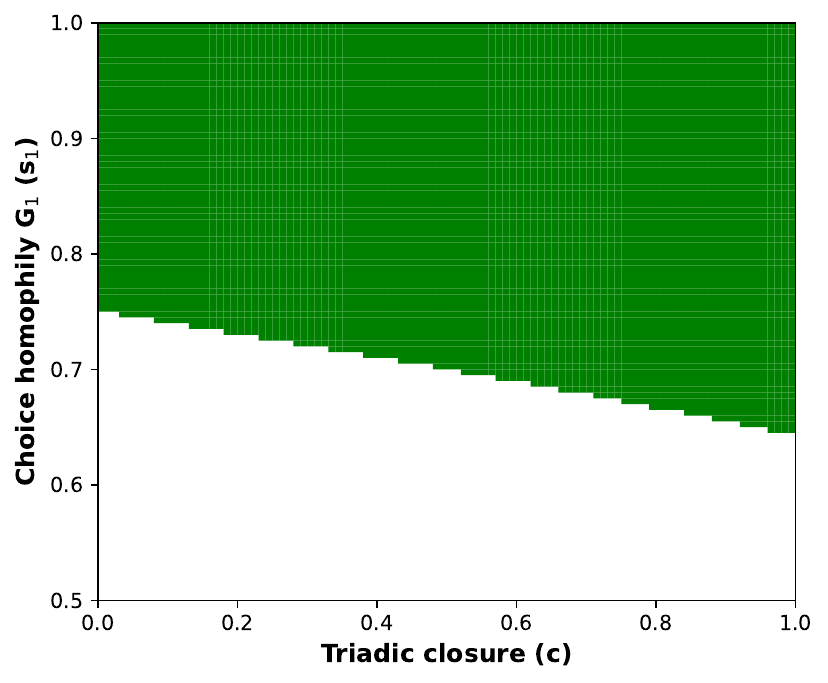}
    \caption{\textbf{Nash equilibrium for symmetric groups.} The cost of expression is $\xi=0.5$. In this phase diagram, the cumulative advantage of the combined effect of the triadic closure and choice homophily at favouirng expression-expression equilibria can clearly be seen.  }
    \label{fig:Diagrama}
\end{figure}

Additionally, in these symmetric scenarios, triadic closure and choice homophily have a highly non-linear combined effect. As the triadic closure probability increases, the impact is not gradual but amplified, reinforcing intra-group connectivity and accelerating the segregation of the network. In turn, the increase of the triadic closure probability always lead to the a scenario in which both groups are able to express themselves simultaneously. This effect is better seen in figure \ref{fig:Diagrama}

Furthermore, promoting the formation of new links through triadic closure appears to be an effective mechanism for helping the minority group overcome the spiral of silence. Because of its greater size, the majority naturally finds itself in a supporting environment in the absence of triadic closure. As a result, majority members are more likely to express themselves, reinforcing their visibility in the network. This increase the peer-pressure over the minority as its willing to speak out is diminished not only by structural constrains but also by the other group expression. However, the structural constrains imposed by triadic closure increases the chances of a minority agent to be surrounded by a locally supportive cluster, improving their social reinforcement and reducing their exposure to the other opinion.
This has the dual effect of lowering the fear of isolation that feeds the spiral of silence and increasing the perceived safety of voicing minority opinions. This shows that encouraging local clustering in social networks could be a good way to maintain diversity of opinion and keep minority viewpoints from being marginalized.

However, increasing the probability of triadic closure must be handled with care. Triadic closure do not only promotes the creation of edges inside the disfavoured group, it also does the same within the already dominant one. Consequently, an excessive clustered driven dynamics often unintentionally deepen the asymmetries, reinforcing the visibility of the majority while further isolating the minority. In extreme cases, this dynamic can ultimately lead to the emergence of core–periphery hierarchies, amplifying structural dominance and limiting the disadvantaged group’s ability to gain expressive influence.

This issue becomes particularly relevant when the two groups exhibit different choice homophilies. In symmetric or low triadic closure scenarios, new expressive phases typically emerge smoothly: a group that is initially silent becomes expressive when the other group remains silent, and eventually both groups express simultaneously. However, when triadic closure is high, even a small change in one group’s homophily bias can produce fundamentally different equilibrium outcomes, such as a sudden shift from mutual expression to a configuration where only the majority group continues to express.

\begin{figure}
    \centering
    \includegraphics[width=0.8\linewidth]{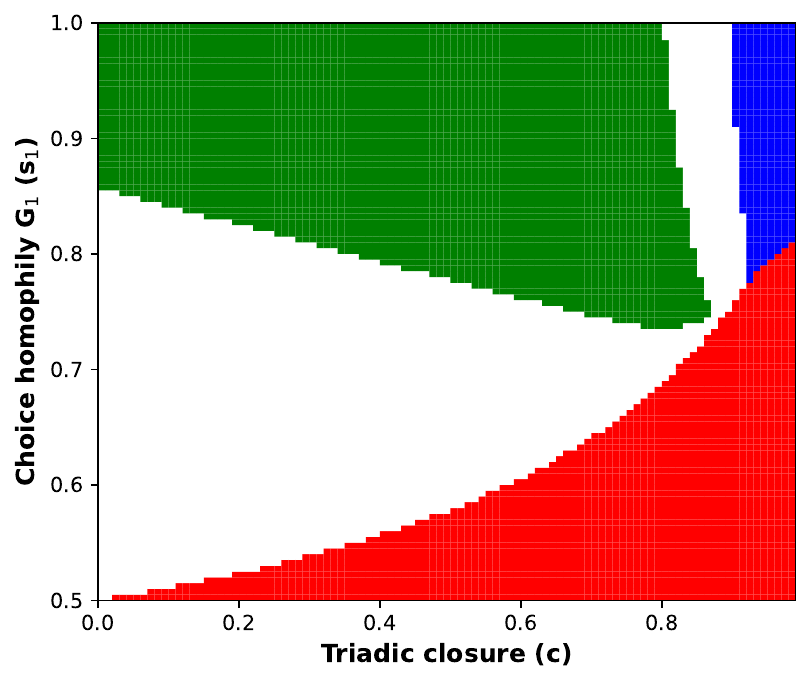}
    \caption{\textbf{Network memory of Nash equilibria}. Different initial conditions in the network gives different shapes for the phase space of Nash equilibria. This figure has been done with the same parameters that the figure unequal groups-different bias of figure \ref{fig:Nasheq} but starting from a different network. The difference between both phase diagram highlight the importance of path dependence in the network formation.}
    \label{fig:memory}
\end{figure}

Although these are relevant results, it is important to remember that the plots in figure \ref{fig:Nasheq} have been obtained from an initial random configuration and that the network evolution dynamics show  path dependence and structural memory. As a consequence, the phase diagram itself is not universal: different initial conditions can lead to qualitatively different equilibrium outcomes, even under the same final parameter values (figure (\ref{fig:memory})). Therefore, interventions with the goal of promoting a concrete public discourse state, have to take into account the current network state.
\subsection{Q-learning}\label{subsection:Q-learning}
Building on the Q-learning framework introduced in Subsection \ref{subsect:Q}, in this subsection we analyze how the dynamics of opinion expression are shaped by two key factors: path dependence and decision stochasticity. First, we explore how varying the structural parameters in forward and backward directions produces hysteresis cycles and a tendency to become locked in one-group expressive states. Second, we examine the effect of a finite temperature, which introduces behavioural noise into the learning dynamics, which allows agents to occasionally explore less optimal strategies and enables them to scape from the trapping states. Together, these analyses highlight the limitations of deterministic equilibrium-based predictions and reveal the importance of memory and randomness in shaping long-term expressive outcomes.

\begin{figure}[b]
    \centering
    \includegraphics[width=0.8\linewidth]{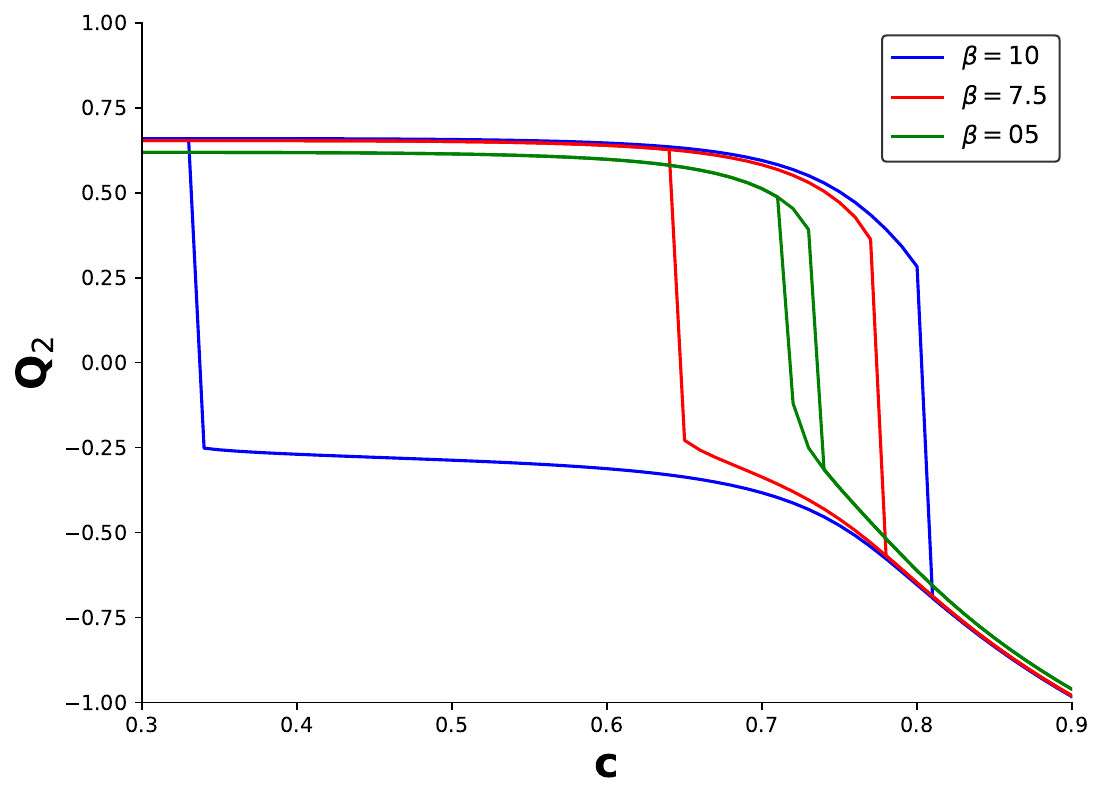}
    \caption{\textbf{Hysteresis cycle in expression dynamics.} The parameters chosen are $n_1=0.2$, $s_{11} = 0.8$ and $s_{22} = 0.6$.  The system alternates between the expression of both groups as the high asymmetry in the choice homophily prevent the system to create big homophilic clusters.}
    \label{fig:HisteresisNet}
\end{figure}

A representative example of path dependence that arise when we vary the triadic closure probability is illustrated in figure \ref{fig:HisteresisNet}. Due to competing structural asymmetries, the network tends to oscillate between two distinct core–periphery architectures, each of which silences one of the groups. Although at the middle point the network presents an homophily amplification topology, the segregation achieved is often insufficient to allow the disadvantaged group to escape from silence and the system remains trapped in this configuration by path memory until it transitions into the opposite topology.

\begin{figure}[b]
    \centering
    \includegraphics[width=0.8\linewidth]{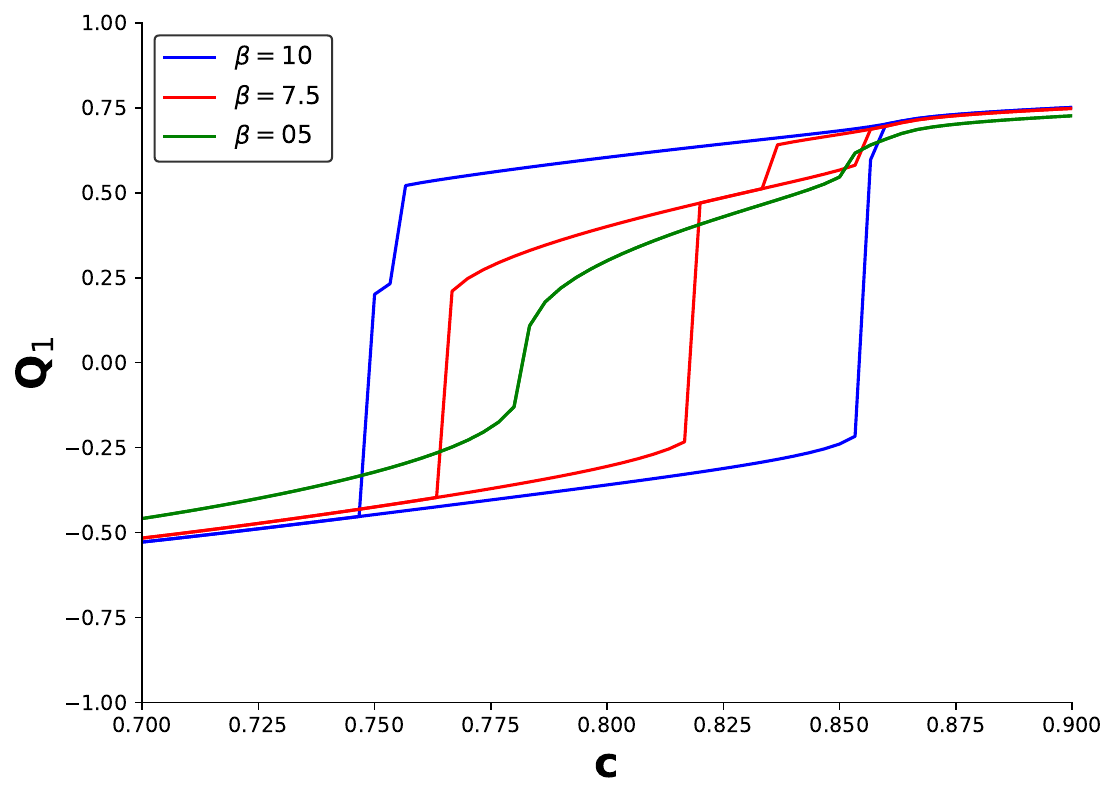}
    \includegraphics[width=0.8\linewidth]{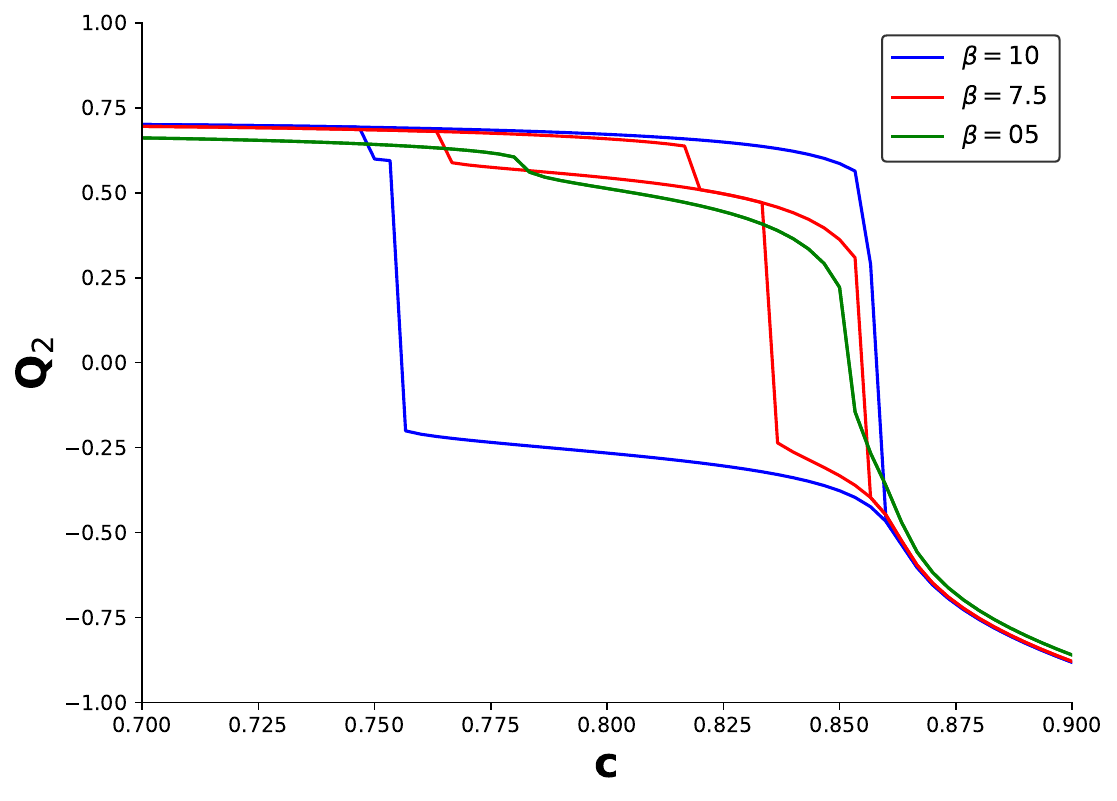}
    \caption{\textbf{Hysteresis cycle in expression dynamics.} The parameters chosen are $n_1=0.2$, $s_{11} = 0.755$ and $s_{22} = 0.65$. For middle values of $c$, the network is lead to an expression-expression state if the temperature is not to low and there are two competing structural biases. }
    \label{fig:HisteresisNet2}
\end{figure}

In contrast, when the structural asymmetries are moderate and the triadic closure probability reaches an intermediate value, the network can evolve toward a configuration in which both groups form highly segregated, internally cohesive clusters. In this scenario, the memory embedded in the Q-learning dynamics ensures that at least one group remains expressive while the other eventually overcome the spiral of silence (figure \ref{fig:HisteresisNet2}). This scape is highly temperature-dependent: if the temperature is too low, the silent group is not able to explore other strategies and will remain trapped in silence. 

However, if the social dynamics does not present such compensating asymetries, the system is unlike to scape the spiral of silence. Whenever a group is benefited both by group size and a higher choice homophily, it will consolidate its dominant role easily for any triadic closure probability. The disadvantaged group, lacking both size support and structural reinforcement, remains silent throughout. Even with moderate stochasticity, the absence of structural counterweights prevents the emergence of balanced expression. This highlights the crucial role of competing asymmetries in enabling expression diversity within structurally biased networks.

\begin{figure}
    \centering
    \includegraphics[width=\linewidth]{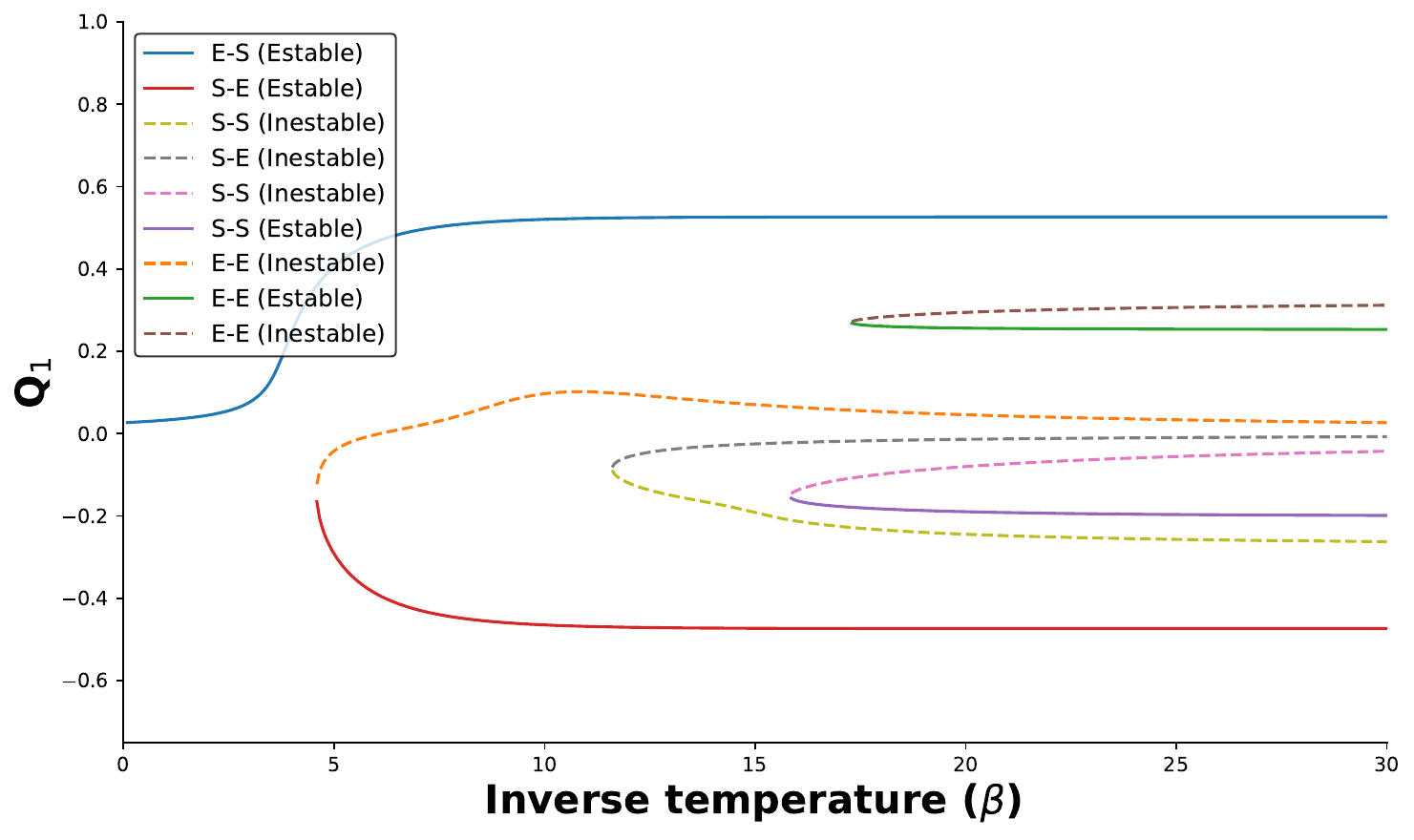}
    \includegraphics[width=\linewidth]{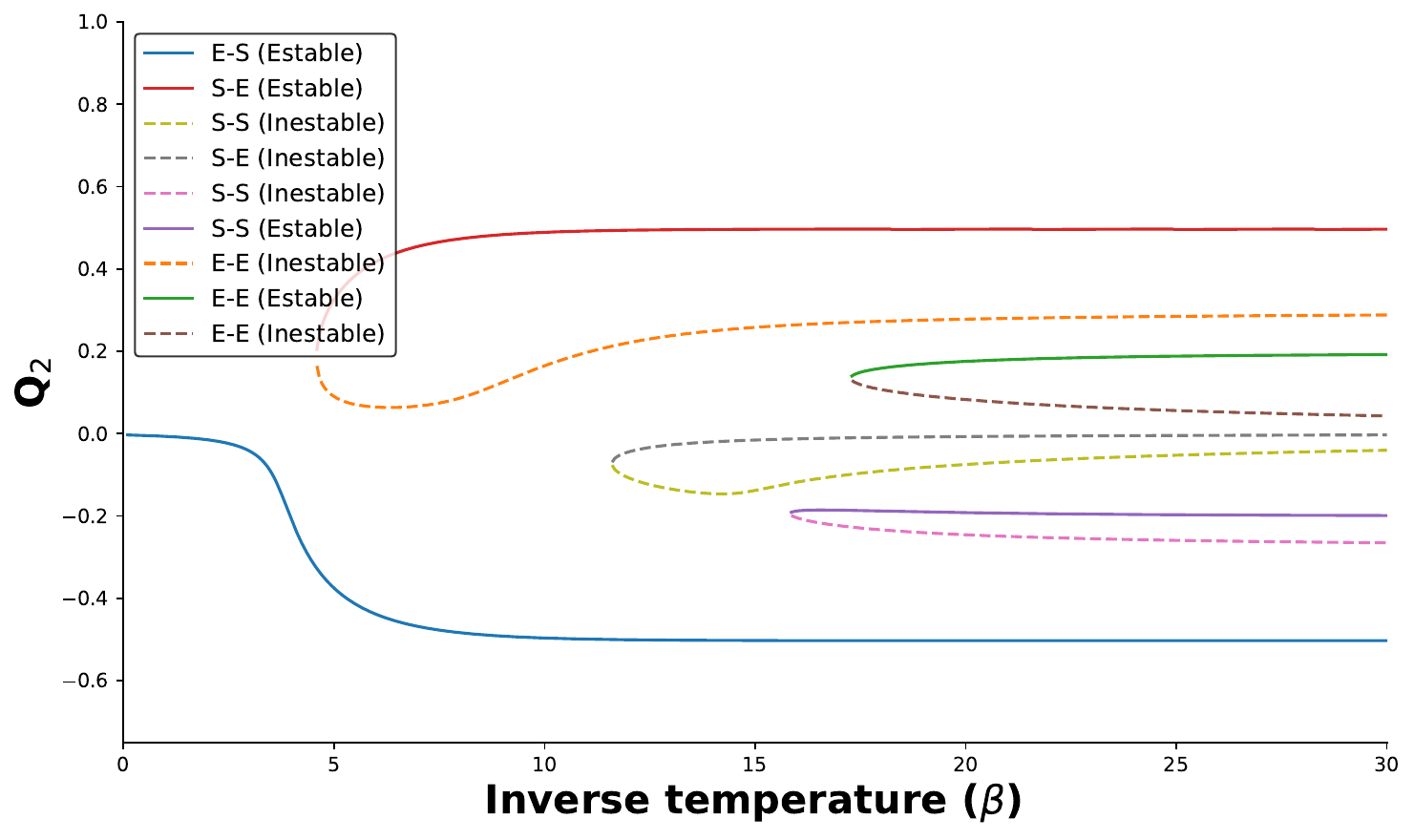}
    \caption{\textbf{Noise transition in homophilic-amplification networks.} The parameters choosen are $s_{11} = s_{22} = 0.65$, $n_1=0.5$, $c=0.785$, $\xi =0.2$. }
    \label{fig:Thermal_transitions}
\end{figure}

Figure \ref{fig:Thermal_transitions} displays the typical behaviour of a population with the temperature parameter. In non-extreme situations, the networks generated exhibit homophilic amplification architectures where both groups maintain a significant number of connections between them. As a consequence, the most stable configuration is typically one in which only one of the groups expresses, while the other remains silent. These asymmetric expressive states are less sensitive to the noise introduced by non-rational agents and are therefore more likely to persist at high temperatures, where decision-making becomes increasingly stochastic.

The other states, on the other hand, are less stable and more sensitive to fluctuations. Configurations in which both groups express simultaneously or neither does require a higher coordination in agent expression. As a consequence, they are difficult to arise under high-temperature dynamics, where stochastic behaviour dominates. Therefore, they appear at lowering the temperature where agents behave more deterministically and are able to coordinate in more complex expression patterns.

As a consequence, the expressive–expressive fixed point is not always reachable in practice. As seen, it usually need low temperatures to reduce the behavioural noise. Conversely, when the temperature is too low, the lack of exploration prevent the system from escaping trajectories caused from initial asymmetries.This trade-off emphasizes the existence of a critical temperature range in which exploration and coordination are both adequate to reach the wanted expressive–expressive state.

\section{Monte Carlo simulations}\label{sec:MonteCarlo}
To complement the mean-field analyses, in this last section we present Monte Carlo agent-based simulations. These simulations allow us to asses the validations of the mean-field theories under stochastic fluctuations and to capture new features that arises from topological features not covered in the mean-field theory such as the clustering correlations and heavy-tails in the degree distribution.

\subsection{Finite-Size effects and dispersion around mean-field prediction}
\begin{figure}
    \centering
    \includegraphics[width=\linewidth]{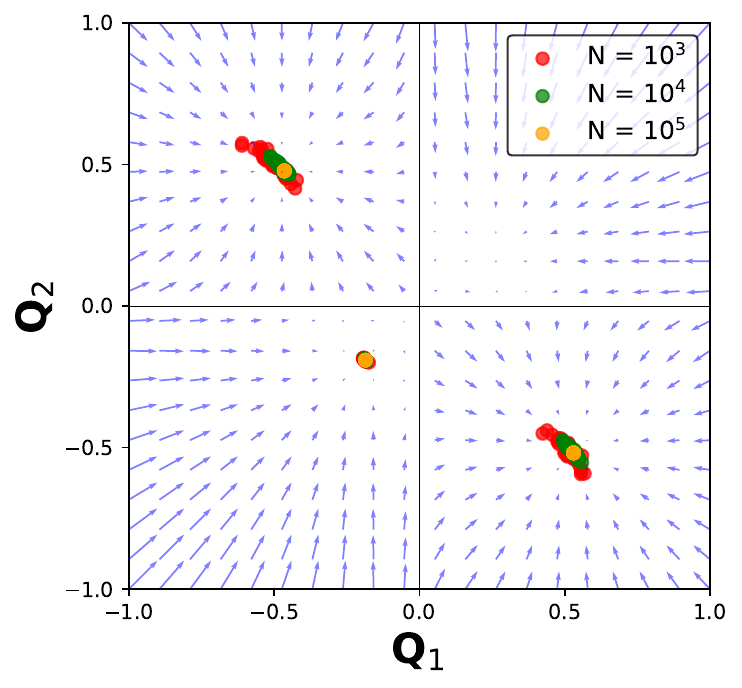}
    \includegraphics[width=\linewidth]{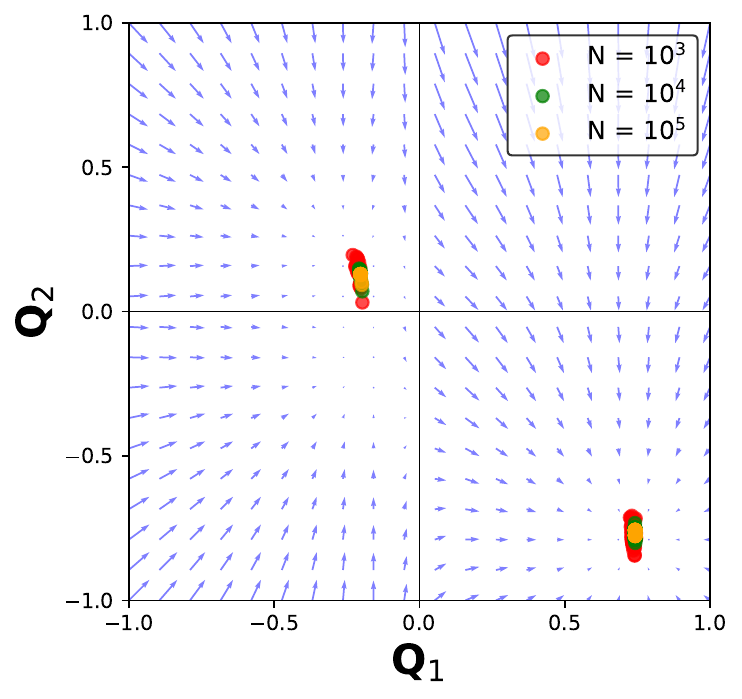}
    \caption{Finite size effect in the convergence to the fixed point. $100$ simulation were performed for each system size each in a different network (except for $N=10^5$). As the system size grows, there is less dispersion around the predicted stable configurations. Additionally, in the real system, some fixed points are not reached. (\textbf{Top}) Homophily amplification network ($n_1=0.5$, $s_{11}=0.6525$, $s_{22}=0.65$, $c=0.785$) at $\beta=20$, $\xi=0.2$. (\textbf{Bottom}) Core-periphery network ($n_1=0.5$,$s_{11}=0.7825$, $s_{22}=0.65$, $c=0.765$) at $\beta=20$, $\xi=0.2$}
    \label{fig:Size}
\end{figure}

To evaluate the robustness of the mean-field predictions in finite systems, we perform Monte Carlo simulations of the Q-learning dynamics over networks with different sizes. Figure \ref{fig:Size} shows the resulting expression states for 100 independent simulations for each of three network sizes and the mean field prediction for the system flow. This study allows us to assess how the convergence behaviour of the stochastic system aligns with the deterministic trajectories predicted by the analytical model.

As can be seen in the figure, finite size has a clear effect about the dispersion around the mean-field prediction. Networks with a smaller number of nodes present a higher dispersion, specially around those fixed points in which one or both groups speaks out. As these points are related with the network structure, this dispersion arises from defects in the network topology. As the number of nodes decreases, the mean-field approximation of the network becomes less accurate, and the predicted fixed points no longer behave as true attractors, but rather as trapping regions around the mean-field values.

\begin{figure}
    \centering
    \includegraphics[width=\linewidth]{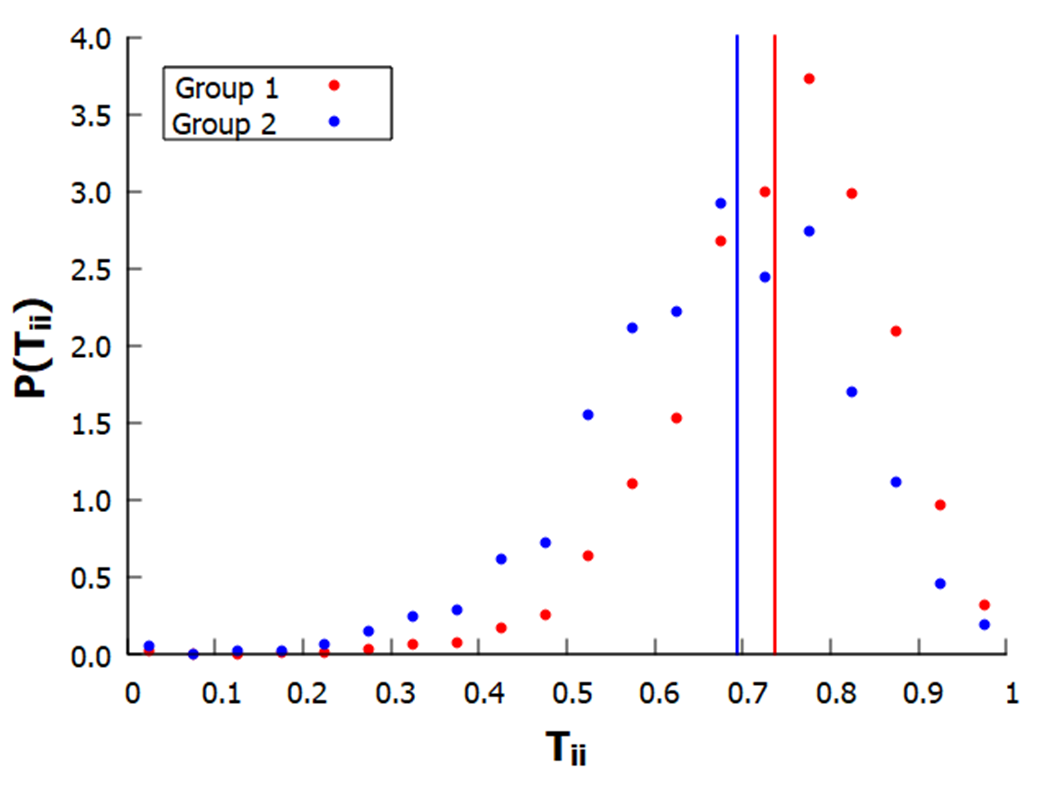}
    \caption{Dispersion of the transition matrix elements around the mean value predicted by the mean-field theory (solid lines). The network has an homophily amplification topology with $n_1=0.5$, $s_{11}=0.6525$, $s_{22}=0.65$, $c=0.785$.}
    \label{fig:Dis}
\end{figure}

Another important observation is that some of the fixed points predicted by the mean-field theory do not appear in the simulations. This discrepancy has two main origins. First, the mean-field network model fails to accurately estimate the $T_ {ij}$ elements when the choice homophily differs between the two groups \cite{asikainen2020cumulative}. Second, even when the predicted mean-field values match the averages observed in the simulated networks, the actual $T_{ij}$ values exhibit high variance across agents (figure \ref{fig:Dis}), introducing significant structural noise into the system. Much like behavioural noise at high temperatures, this structural variability tends to disrupt the more fragile fixed points s-s and e-e, as they require a higher coordination. This effect is especially pronounced at networks created with a high triadic closure probability and, more precisely in the periphery of core-periphery structures.

\subsection{Structural and Thermal Phase Transitions}
In this subsection, we explore how the different opinion expression phases arise in the agent-base model under changes either in behavioural noise and network evolution parameters. Specifically, we simulate two different experiments: one in which the temperature is varied while fixing the other social dynamics parameters and another where a random network evolves with different triadic closure probabilities and a fixed temperature.

\begin{figure}[h]
    \centering
    \includegraphics[width=\linewidth]{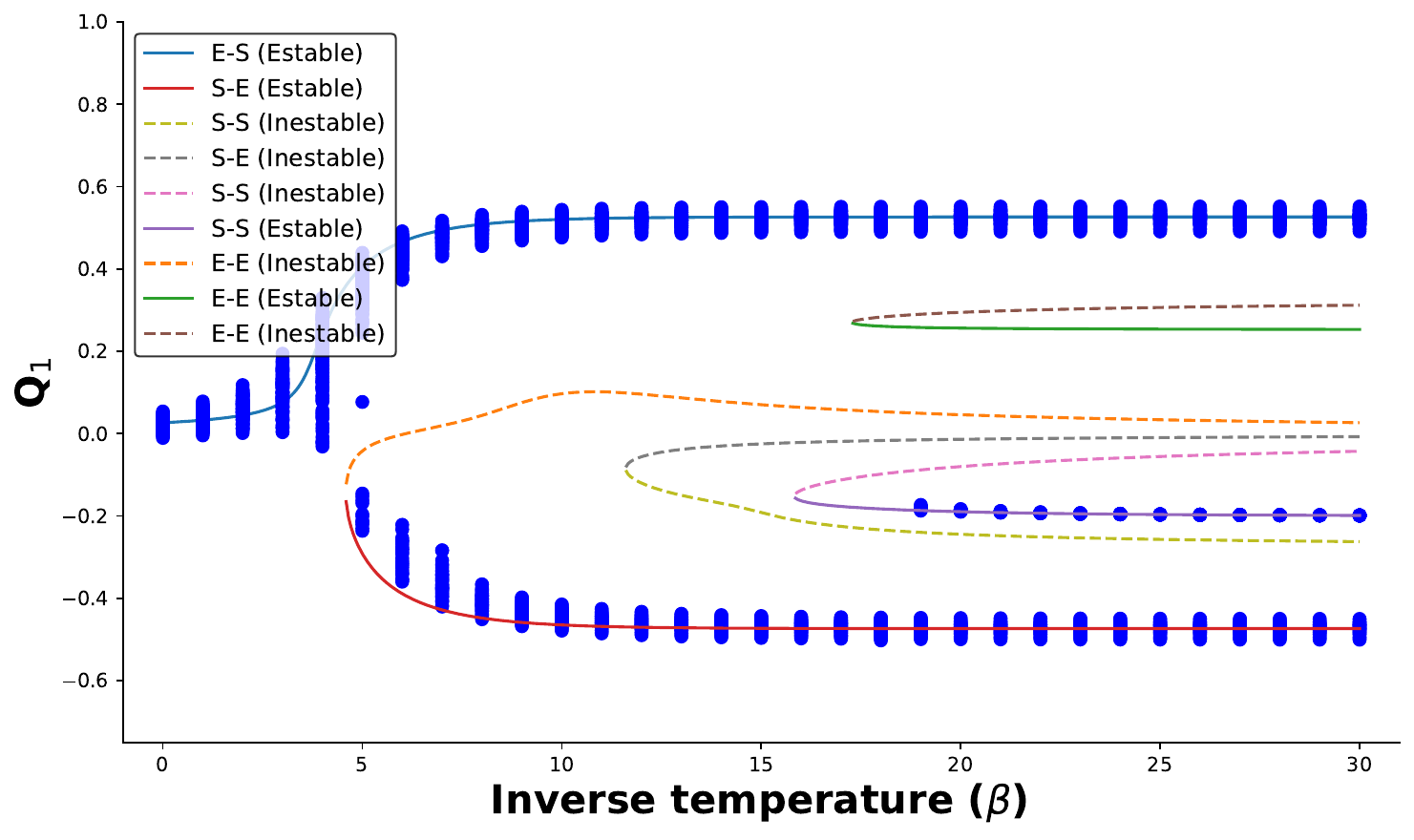}
    
    \includegraphics[width=\linewidth]{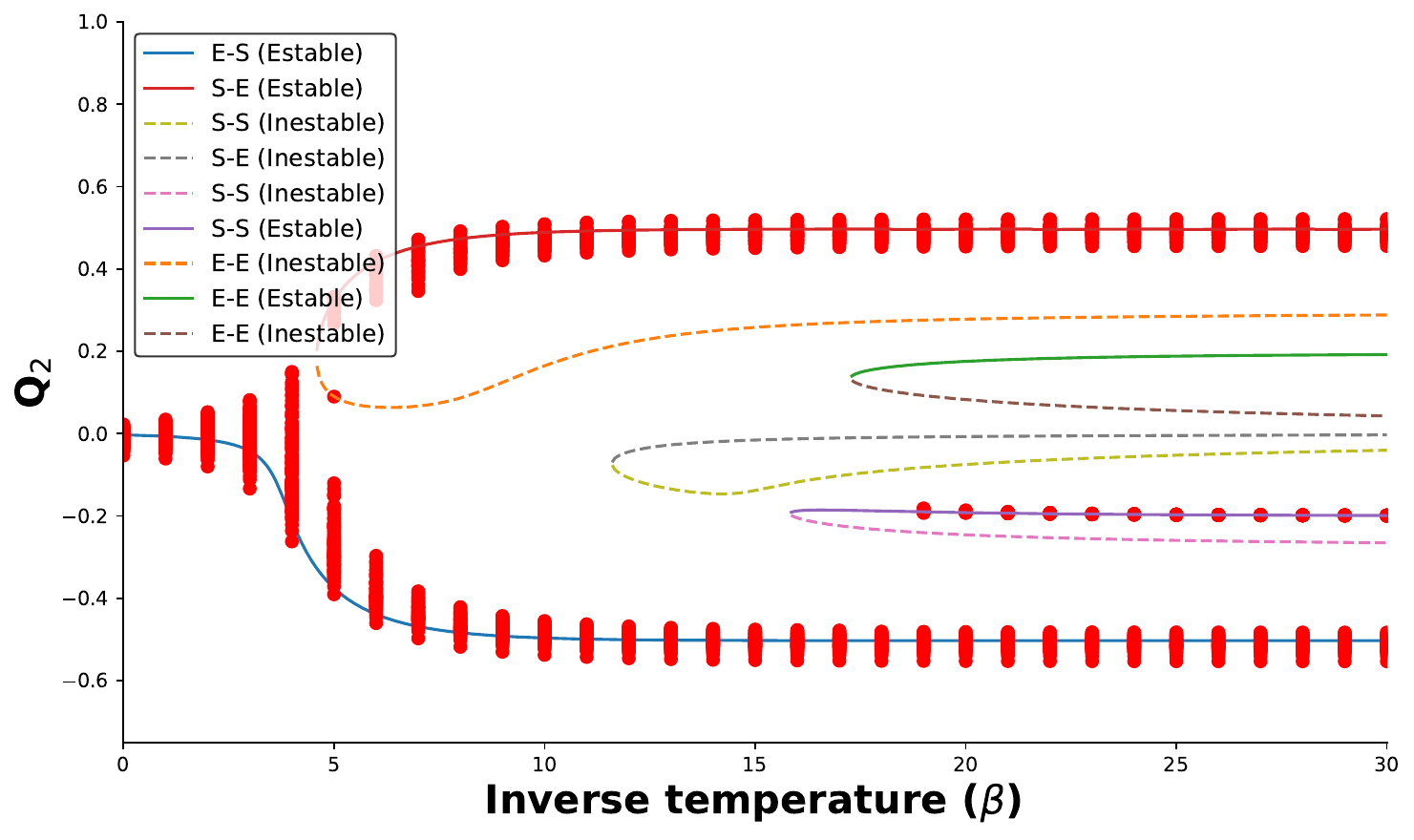}
    \caption{\textbf{Stationary expression states as a function of the inverse temperature $\beta$.} Solid lines represent the fixed points predicted by the mean-field approximation, while the points indicate the results of Monte Carlo simulations over finite networks. The simulations were carried with $N=10^4$ nodes, equal size groups, $c=0.785$, $s_{11}= 0.6525$, $s_{22} = 0.65$ and $\xi=0.2$. A general agreement between theoretical predictions and simulation results is observed, although the E–E state and transition misprediction highlight the influence of stochasticity.}
    \label{fig:SimulacionTemperatura}
\end{figure}

In the first set of simulations, we fix the network structure and instead vary the inverse temperature. As shown in figure \ref{fig:SimulacionTemperatura}, we observe a general good agreement between the mean-field prediction and the simulated system. However, it also shows that the role of noise in the real system can not be underestimated. First, as discussed in the previous subsection, the expression–expression phase fails to appear in the simulations as a consequence of the structural noise. Second, noise also affects the system’s behaviour near bifurcation points: at high temperatures, behavioural randomness combined with structural variability leads to greater dispersion in the outcomes; whereas at low temperatures, structural irregularities delay the emergence of coordinated states, shifting the transition thresholds with respect to theoretical predictions.

\begin{figure}
    \centering
    \includegraphics[width=\linewidth]{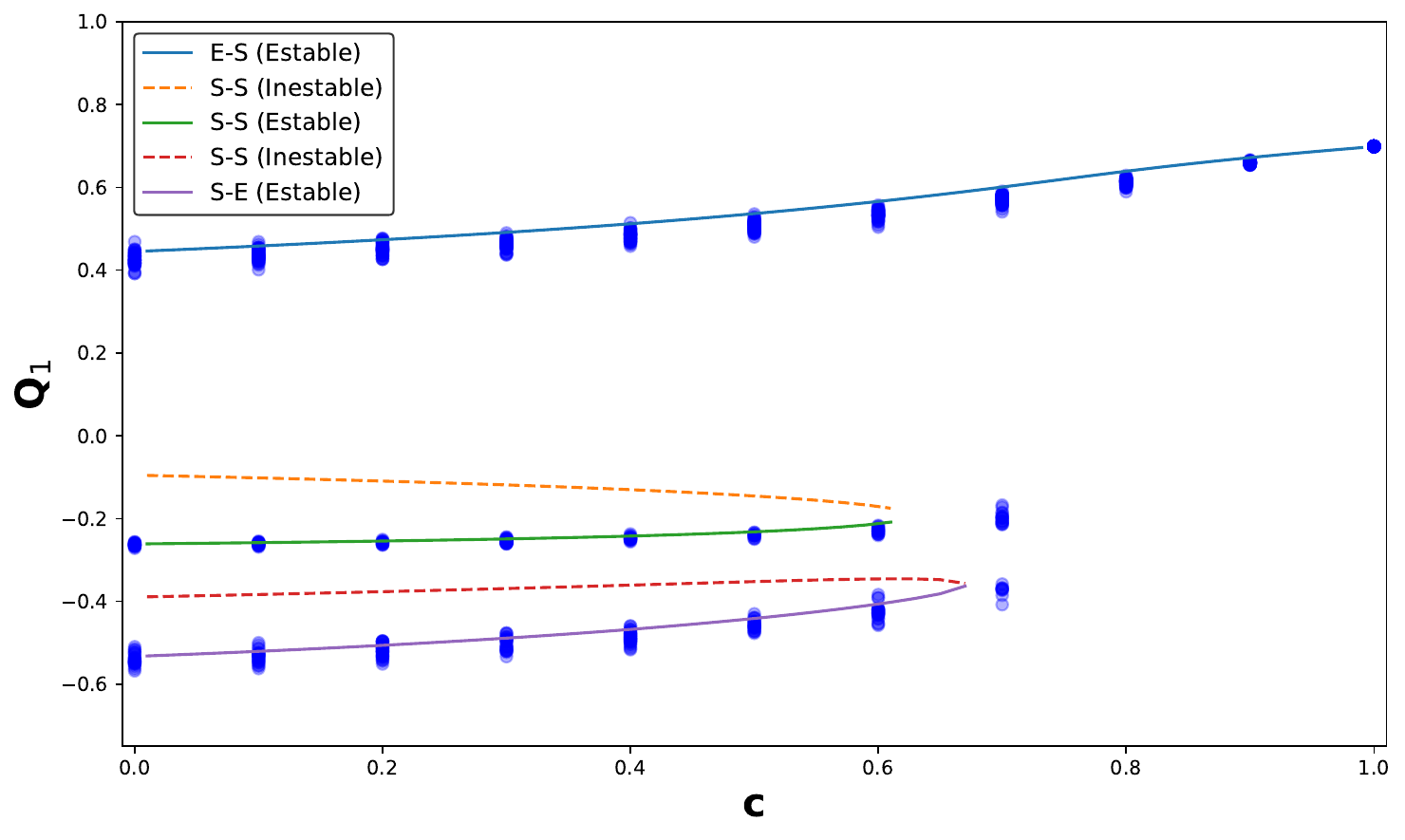}
    \includegraphics[width=\linewidth]{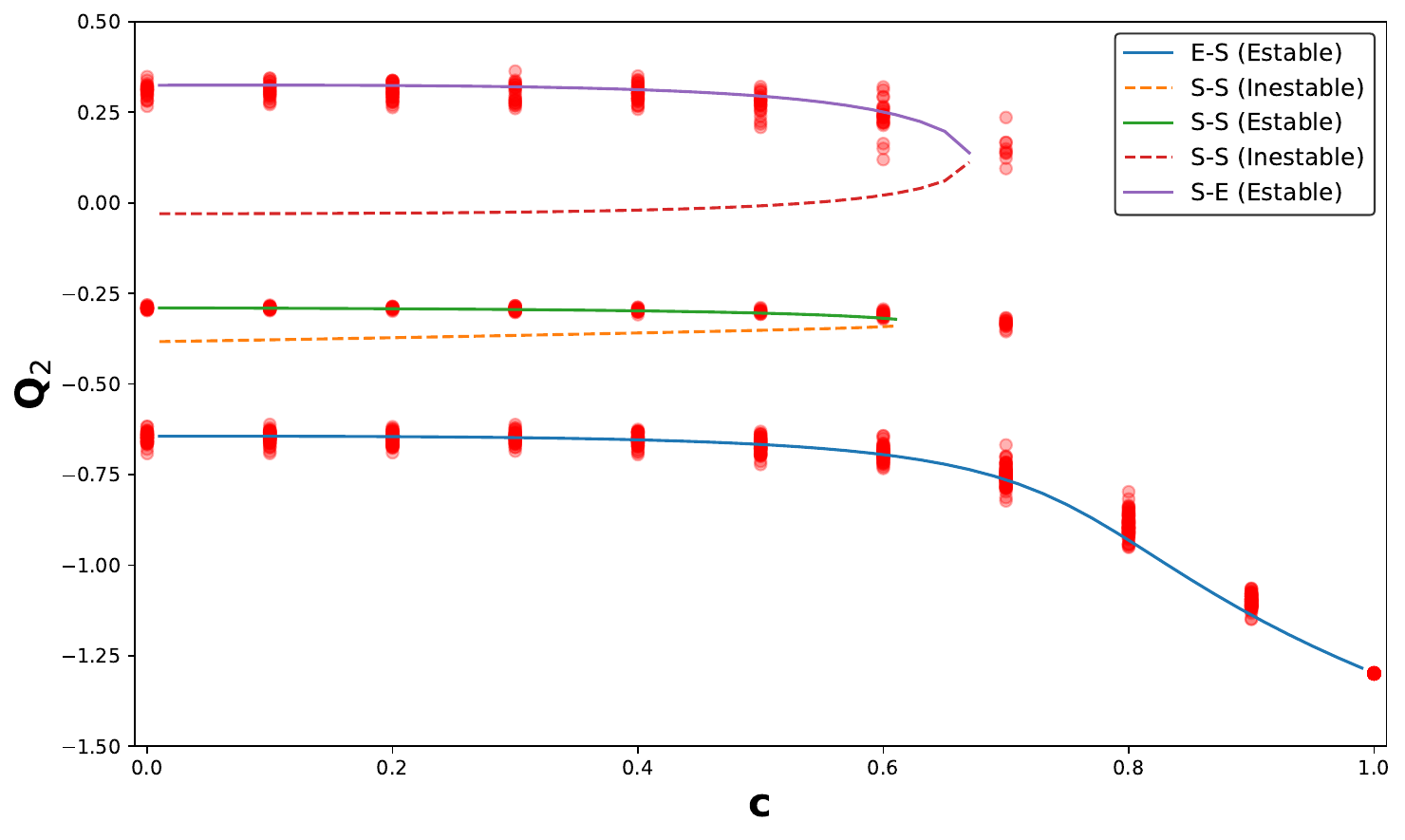}
    \caption{\textbf{Mean-field bifurcation diagram as a function of the triadic closure probability $c$.} The coloured curves represent different stable and unstable fixed points predicted by the mean-field dynamics. The simulations were carried over networks with $N=10^3$ agents, equally distributed in both groups, $s_{11} = 0.755$, $s_{22} =0.65$, $\beta=10$ and $\xi=0.2$.}
    \label{fig:Structural_simulations}
\end{figure}

In the second experiment, we fix the temperature and progressively vary the triadic closure probability $c$. Here, the network evolves for each $c$ value independently, so there is no structural memory from other $c$ values networks. The results are shown in figure \ref{fig:Structural_simulations}. The results confirm the qualitative behaviour predicted by the mean-field model: the system exhibits saddle transitions between different expressive regimes and includes regions of multistability. There are some small deviations at the critical values where the transitions occur. However, in this case, they appear to be caused not by structural noise, but rather by a slowdown in the dynamics near the bifurcation points, which delays the convergence to the stationary state during the simulations.

\subsection{Hysteresis Cycles in stochastic Simulations}
We now turn to simulate the hysteresis cycles previously identified in the mean-field theory to test if these also emerge in the stochastic system. In this subsection, as previously done, we reproduce a hysteresis experiment by varying the triadic closure probability in both directions while keeping the other model parameters constant. The network is updated from the previous at each time step until it stabilizes, and the expression dynamics is allowed to converge then. This procedure allows us to analyze whether finite-size stochasticity preserves, distorts, or removes the hysteresis observed in the mean-field approximation.
\begin{figure}
    \centering
    \includegraphics[width=0.9\linewidth]{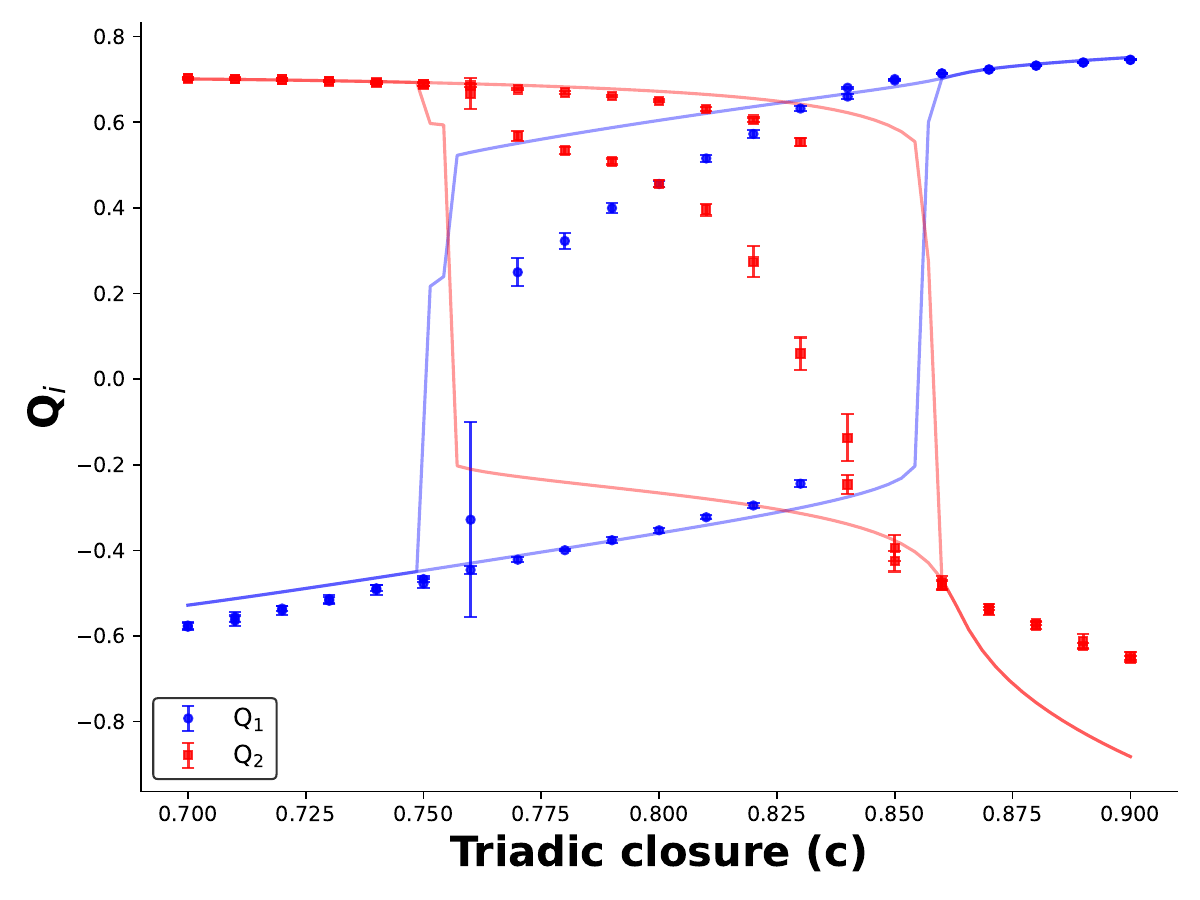}
    \caption{\textbf{Hysteresis experiment simulation.} A society is changed quasi-statically by changing the triadic closure probability. Parameters of the simulations are $n_1=0.2$, $s_{11} = 0.755$, $s_{22} = 0.65$, $\beta = 10$, $\xi = 0.2$ and $N=10^4$.}
    \label{fig:Hysteresis-exp}
\end{figure}

A comparison between the simulated experiment and the mean-field result is shown in figure \ref{fig:Hysteresis-exp}. We observe that hysteresis is indeed present in the stochastic system. However, the loop is narrower and less pronounced compared to the mean-field prediction. This behaviour can be explained by the structural noise introduced in core-periphery architectures, where irregular topologies cause certain nodes to deviate from the predicted state by the mean-field model. These local deviations act as early activators, facilitating the system's escape from metastable states and reducing the persistence of the current state.

\section{Conclusions}\label{sec:conclusion}

In this work, we have explored the role of the mechanisms driving the development of social networks in shaping the patterns of public speech that emerge within them. To do so, we integrated two complementary models: a network evolution model that combines triadic closure and choice homophily, and an opinion expression model based on game-theoretic reasoning and dynamic Q-learning.

By combining the mean-field equations for network formation with the Nash equilibrium conditions from the expression model, we have studied the conditions under which different expressive states can arise as stable solutions of the system. This analysis revealed the central role of triadic closure as an amplifier of not only group homophily but also differences between groups. In particular, we found that a moderate level of triadic closure can promote expression across both groups by reinforcing internal cohesion. However, at higher levels, the same mechanism can lead to silence of structurally disadvantaged groups, especially when combined with asymmetries in choice homophily.

Through the bifurcation analysis of the Q-learning dynamics we demonstrated how hysteresis and path dependence naturally emerge, producing situations where once a group has fallen into silence, recovery becomes difficult even when conditions shift. Specifically, we find that the moderate levels of triadic closure required to sustain expression for both opinion groups typically do not lead to extreme polarization in the network and therefore they do not guarantee spontaneous transitions away from silenced configurations.This suggests the existence of a delicate balance between exploration and coordination: sufficient behavioural noise is needed to escape previously stable silent states, while enough determinism is required to let the system converge toward mutually expressive coordinated outcomes. These results underline the importance of dynamic adaptation and learning in shaping long-term expressive behaviour, beyond what can be captured by static equilibria.

Monte Carlo simulations have confirmed several of the theoretical predictions derived from the mean-field and Q-learning analyses. However, they have also highlighted the need for a better understanding of the role of structural noise, which can act as a destabilizing force, particularly for mutual expression or mutual silence states. These results underscore the importance of accounting for irregularities when applying theoretical models to finite and heterogeneous populations.

Taken together, these results highlight a delicate balance: moderate levels of clustering can sustain diversity of expression by protecting minority groups, while excessive clustering can marginalize them and reinforce structural dominance. This finding has direct implications for online platforms, where recommendation algorithms that mimic triadic closure may either sustain or suppress minority voices depending on parameter regimes. Looking forward, extending the present framework to include richer behavioural strategies, such as opinion change, misrepresentation, or adaptive avoidance of hostile environments, will bring the model closer to real-world discourse dynamics. Empirical validation on social media or survey data is a natural next step, providing a bridge between our theoretical predictions and the complex realities of public debate.

\begin{acknowledgments}
E.C. acknowledges support from the Spanish grants PID2021-128005NB-C22, funded by MCIN/AEL 10.13039/501100011033, and from Generalitat de Catalunya under project 2021-SGR-00856.
\end{acknowledgments}

\section*{Appendix I: Q-Learning in low-temperature}
This appendix is developed to explore the low temperature regime of the dynamical model presented in subsection \ref{subsect:Q}. This limit is interesting to better understand the connection between the Nash-equlibria framework and the Q-learning dynamics.

At low temperatures, the probability of expression becomes deterministic. This is reflected in the convergence of the sigmoid function to the Heavyside function:
\begin{equation}
    \frac{1}{1+e^{-\beta Q}} \to \Theta(Q)
\end{equation}
With that change, the dynamical equations can now be written as:
\begin{equation}
    \frac{dQ_1}{dt} = T_{11}\Theta(Q_1) - (1-T_{11}) \Theta(Q_2) -Q_1-\xi
\end{equation}
\begin{equation}
    \frac{dQ_2}{dt} = T_{22}\Theta(Q_2) - (1-T_{22}) \Theta(Q_1) -Q_2-\xi
\end{equation}
The jacobian matrix, needed to obtain the stability of the fixed points can be written as:
\begin{equation}
    J = \begin{pmatrix}
        -1 + T_{11}\delta(Q_1) & (1-T_{11})\delta(Q_2)\\
        (1-T_{22})\delta(Q_1) & -1 + T_{22}\delta(Q_2)
    \end{pmatrix}
\end{equation}
Note that, whenever $Q_1^*\neq 0$ and $Q_2^* \neq0$, the fixed points are always stable.

The system can now be solved in each plane's quadrant and then see the conditions that must be fullfilled in order to that fixed point really exist. For example, the solution in the $Q_1>0$, $Q_2>0$ quadrant, that correspond to the expression-expression equilibrium, is
\begin{equation}
    (Q_1^*, Q_2^*) = (2T_{11} - 1-\xi, 2T_{22}-1-\xi)
\end{equation}
However, we have supposed that both $Q_1^*, Q_2^* >0$. Thus, in order to exist this fixed point, 
\begin{equation}
    T_{11} > \frac{1+\xi}{2}
\end{equation}
and the same for $T_{22}$. These are the same conditions obtained through the game theory formulation. As a consequence, both game theory and Q-learning framework are equivalent in the $\beta \to \infty$ limit.

Those stable fixed points are disconnected through saddle points that appear near to the axis. In order to obtain the position of those saddle points, it is necessary to take again the limit $\beta \to \infty $ but keeping one of the product $\beta Q_i$ small. This will allow us to take for one group the approximation $p_j \to \Theta(Q_j)$ while for the other we can perform a Taylor expansion.

For example, for the saddle point between the silence-silence and expression-silence fixed point, we take the limit $p_2 \to 0$ (as $\beta \to \infty$ and $Q_2 < 0$) while $p_1 \approx 1/2 + \beta Q_1/4$. By substitution:
\begin{equation}
    \frac{dQ_1}{dt} = T_{11}\left[\frac{1}{2} + \beta \frac{Q_1}{4}\right] - Q_1 -\xi
\end{equation}
\begin{equation}
    \frac{dQ_2}{dt} = (1- T_{22})\left[\frac{1}{2} + \beta \frac{Q_1}{4}\right] - Q_2 -\xi
\end{equation}
The solution for the fixed point then is:
\begin{equation}
    (\beta Q_1^*, \beta Q_2^*) \approx \left(\frac{4\xi}{T_{11}} -2, -\xi \beta \frac{1-T_{22}+T_{11}}{T_{11}}\right)
\end{equation}

Finally, there is an additional fixed point near the origin. This is an unstable fixed point. For this fixed point, it is possible to develop both $p_1$ and $p_2$ as Taylor series of $\beta Q_j$, but it often requires multiple orders to give the correct solutions.
\section*{Appendix II: simulation code}

 All code was developed in the C programming language, compiled as a shared library, and executed in parallel using a Python script via the \textit{multiprocessing} library. The code can be found in \href{https://github.com/JuanCastillo29/Spiral-of-Silence-in-Clustered-Homophilic-Networks.git}{https://github.com/JuanCastillo29/Spiral-of-Silence-in-Clustered-Homophilic-Networks.git}.

\bibliographystyle{IEEEtran}
\bibliography{Bibliography}

\end{document}